\documentclass[12pt,preprint]{aastex}
\usepackage[yyyymmdd,hhmmss]{datetime}

\shorttitle{WRs in the Magellanic Clouds}

\shortauthors{Massey et al.}

\begin{document}

\title{A Modern Search for Wolf-Rayet Stars in the Magellanic Clouds. II. A Second Year of Discoveries\altaffilmark{1}}

\author{Philip Massey\altaffilmark{2,3}, Kathryn F. Neugent\altaffilmark{2}, and Nidia Morrell\altaffilmark{4}}

\altaffiltext{1}{This paper includes data gathered with the 1 m Swope and 6.5 m Magellan Telescopes located at Las Campanas Observatory, Chile.}
\altaffiltext{2}{Lowell Observatory, 1400 W Mars Hill Road, Flagstaff, AZ 86001; phil.massey@lowell.edu; \\kneugent@lowell.edu.}
\altaffiltext{3}{Also Department of Physics and Astronomy, Northern Arizona University, Box 6010, Flagstaff, AZ 86011-6010.}
\altaffiltext{4}{Las Campanas Observatory, Carnegie Observatories, Casilla 601, La Serena, Chile; nmorrell@lco.cl.}
%\altaffiltext{4}{Department of Physics and Astronomy \& Pittsburgh Particle Physics, Astrophysics, and Cosmology Center (PITT PACC), University of Pittsburgh, Pittsburgh, PA 15260; hillier@pitt.edu.}

\begin{abstract}

The numbers and types of evolved massive stars found in nearby
galaxies provide an exacting test of stellar evolution models.
Because of their proximity and rich massive star
populations, the Magellanic Clouds have long served as the linchpins
for such studies.  Yet the continued accidental discoveries of
Wolf-Rayet (WR) stars in these systems demonstrate that our knowledge
is not as complete as usually assumed.  Therefore, we undertook a
multi-year survey for WRs in the Magellanic Clouds.
Our results from our first year (reported previously) confirmed nine
new LMC WRs. Of these, six were
of a type never before recognized, with WN3-type emission
combined with O3-type absorption
features.  Yet these stars are  2-3 magnitudes too faint
to be WN3+O3~V binaries.  Here we report on the second
year of our survey, including the discovery of four more WRs, two
of which are also WN3/O3s, plus two ``slash" WRs.  This brings the total of LMC WRs known to 152, 13 (8.2\%) of which were found by our survey, which is now $\sim$60\% complete.
We find that the spatial
distribution of the WN3/O3s is similar to that of other WRs in
the LMC, suggesting that they are descended from the same progenitors.
We call attention to the fact that five of the 12 known SMC
WRs may in fact be similar WN3/O3s rather than the binaries they
have often been assumed to be.  We also discuss our other discoveries:
a newly found Onfp-type star, and a  peculiar emission-line
object. Finally, we consider the completeness limits of our survey.
\end{abstract}


\section{Introduction}
\label{Sec-intro}

For many years our knowledge of the WR population of the Magellanic Clouds was considered ``essentially" complete (see, e.g., Massey \& Johnson 1998).  The accidental discovery of several weak-lined WN-type WRs over the years, and our own discovery of a strong-line WO star (Neugent 2012a),  made us realize that if we wanted an accurate value for the relative numbers of WC-, WO-, and WN-type WRs for comparing with evolutionary models, we needed to conduct a modern survey for WRs using the same techniques we had successfully applied to M33 (Neugent \& Massey 2011) and M31 (Neugent et al.\ 2012b).   We therefore began a multi-year observational program using the Las Campanas Swope 1-m telescope to image fields in the LMC and SMC through a set of interference filters centered on C~III~$\lambda 4650$, He~II~$\lambda 4686$, and neighboring continuum.  Candidates are identified by eye after the frames are reduced and run through  image subtraction software.  Follow-up spectroscopic observations of the WR candidates are then carried out with the Magellan 6.5-m Clay telescope.

The results from the first year of this project were reported by Massey et al.\ (2014, Paper~I), which further describes the motivation  and our observing strategy.  When we began our survey, we realized it was possible that we would find nothing very interesting.  Instead, with the survey only 15\% complete, we had already confirmed nine newly found WRs in the LMC, a 6\% increase in the total number of LMC WRs known.    Of these, one was another very rare WO-type WR, only the third known in that galaxy.  All of the WRs discovered since the Breysacher et al.\ (1999) catalog (BAT99), including those from our survey, were of WN type (with the exception of the two WOs), confirming our suspicions that previous surveys had been biased towards WCs, as these stars have much stronger lines and thus are easier to find (Massey \& Johnson 1998).  

Most interesting, however, was that six of these newly found WN stars were unlike any other WRs previously recognized, showing a WN3-type emission spectrum combined with an O3-type absorption spectrum.  We realized that these were unlikely to be true WN3+O3~V pairs. For one thing, O3~V stars are extremely rare: only a few dozen are known in all of the LMC, and the idea
that six would be paired with newly found WN3s was simply not credible.  It would also be impossible to understand in terms of (single-star) evolution: a very massive star will be in the O3~V stage only for the first million years or so of its life, but it takes 3-4 Myr to produce a WR star.  Finally, and most definitively, these newly found WRs were quite faint, with $V$$\sim$16.  They had absolute visual magnitudes $M_V$$\sim$$-3$, about 2.5~mags fainter than the $M_V$ of O3~V stars.  In other words, these stars were impossibly faint to be WN3+O3~V binaries.  We succeeded in modeling the optical spectrum of one of these stars using CMFGEN (Hillier \& Miller 1998), finding that we could reproduce the absorption {\it and} emission-line spectra with a single set of physical parameters,  using an effective temperature of 100,000~K, a He/H number ratio of 1.0, a nitrogen mass-fraction of 0.011 (corresponding to 10$\times$ solar using the Anders \& Grevesse 1989 values), reduced carbon and oxygen abundances (0.05$\times$ solar), and a mass-loss rate of $1.2\times 10^{-6} M_\odot$ yr$^{-1}$, with a clumping volume filling factor of 0.1. Compared to a recent modeling study of other LMC WNs by Hainich  et al.\ (2014), the 100,000 K effective temperature was at the high end of what was normally found but not anomalous, the bolometric luminosity ($\log L/L_\odot$$\sim$5.6) was at the low end, but also not anomalous, but the mass-loss rate was quite atypical, down a factor of 3-5$\times$ from what we would expect (Massey et al.\ 2015).   We are currently in the process of refining this model based upon recently acquired {\it HST} UV and Magellan NIR data, as well as extending our study to other WN3/O3s.

We have now completed the second year of our survey, including spectroscopic follow-up of our candidates, and report here the discovery of two additional WN3/O3 stars, two newly found WR ``slash stars," plus an odd Of-type star and a peculiar emission-line star.  

%\bigskip
%\bigskip
\bigskip

\section{Observations and Reductions}

As stated above, our interference-filter imaging was conducted using the 1-m Swope telescope. The SITe\#3 CCD camera that we used in the first year of our survey had been replaced with a new system containing a 4110$\times$4096 e2v device with 15$\mu$m pixels.   This provides a  29\farcm8 (EW) $\times$29\farcm7 (NS) field of view, or 0.25 deg$^2$, a $2.6\times$ improvement over the old system.  The chip is read out using four amplifiers, and so the read-time dropped from 127s with the SITe\#3 camera to 37s with the new camera, another substantial improvement.   The e2v camera is a bit more sensitive as well, although we kept to the same exposure times: 300s through each of our three interference filters.  Further details are given in Paper I.

For calibration, we obtained ten bias frames each day, although there was very little ($\leq$1~ADU) bias structure with the new camera. With the large field of view, dome flats contained a 10\% gradient, and we relied upon twilight flats for flat-fielding. We obtained 5 or more twilight sky flats through each filter during the clear nights, dithering the telescope between each exposure. If there was cirrus, we used flats from a clear night.  There is a significant shutter pattern present from the iris shutter in short exposures (1\% peak to valley in a 5s exposure); in order to take advantage of
rapidly fading twilight, we measured the shutter pattern using dome flats during one afternoon, and then used this pattern to correct the short twilight exposures appropriately\footnote{This is a common issue with iris shutters, and as we are unaware of any explicit guidance for dealing with the problem, provide some here. Let $f(x,y)$ be the ``extra" time that an iris shutter is open (in addition to the requested exposure time) as a function of position on the chip.  (Of course, $f(x,y)$ can be negative.)  Assume that one has a stable and constant (voltage-regulated) flat-field system (i.e., dome ``white spot") that illuminates the chip with a pattern S(x,y) in counts/sec.   Then a 1s exposure will produce an image $I_1(x,y) = S(x,y) \times (1+f(x,y))$ above bias.  A 20s exposure will produce an image $I_{20}(x,y)=S(x,y) \times (20+f(x,y))$ above bias.   Dividing one image by the other (after bias removal) will produce $I_1(x,y)/I_{20}(x,y) = (1+f(x,y))/(20+f(x,y))$.  If $f(x,y)$ is a small fraction of a second (either positive or negative), then   $I_1(x,y)/I_{20}(x,y)\approx(1+f(x,y))/20$.  Thus $f(x,y)= 20\times I_1(x,y)/I_{20}(x,y)  - 1$.  Each twilight exposure of exposure time $t$ can be simply corrected (after overscan and bias subtraction) by dividing by $(t+f(x,y))$ before co-adding.}.  Modest corrections for non-linearity were applied to the data from each quadrant using the correction coefficients kindly provided by Carlos Contreras based on measurements made on August 24, 2014; the corrections amount to a maximum of 2.8\%, 1.1\%, 0.8\%, and 1.4\%  for the four quadrants over the full data range.

\begin{figure}
\epsscale{1.0}
\plotone{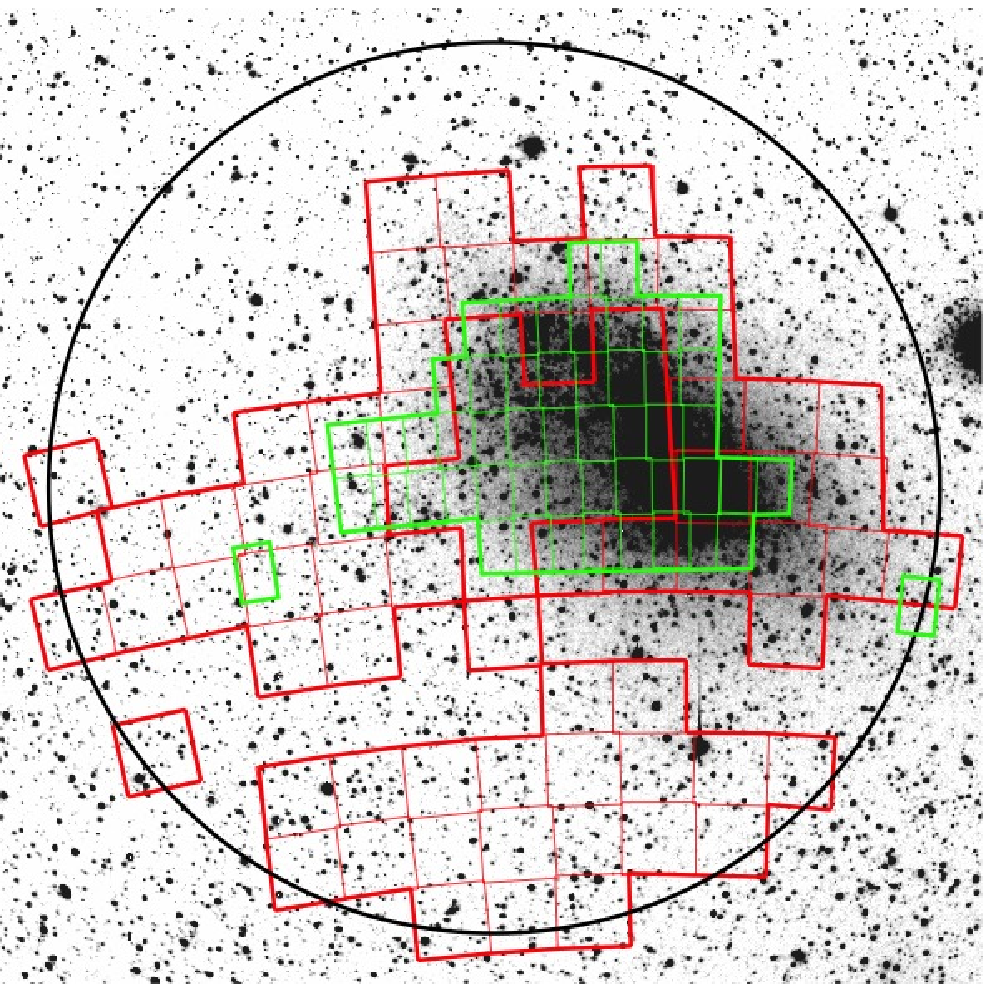}
\caption{\label{fig:SMC} Survey coverage of the SMC.  The green outlines denote the fields we observed in 2013 (Paper I), while the red  denotes the fields covered in 2014 (this paper).  The large circle shows the regions of our intended survey and has a diameter of 6$^\circ$ and is centered on $\alpha_{\rm 2000}$=1$^h$08$^m$00$^s$ $\delta_{\rm 2000}$=$-73^\circ$10\arcmin00\arcsec. The image come from the R-band ``parking lot" camera image of Bothun \& Thompson (1988).}
\end{figure}

\begin{figure}
\epsscale{1.0}
\plotone{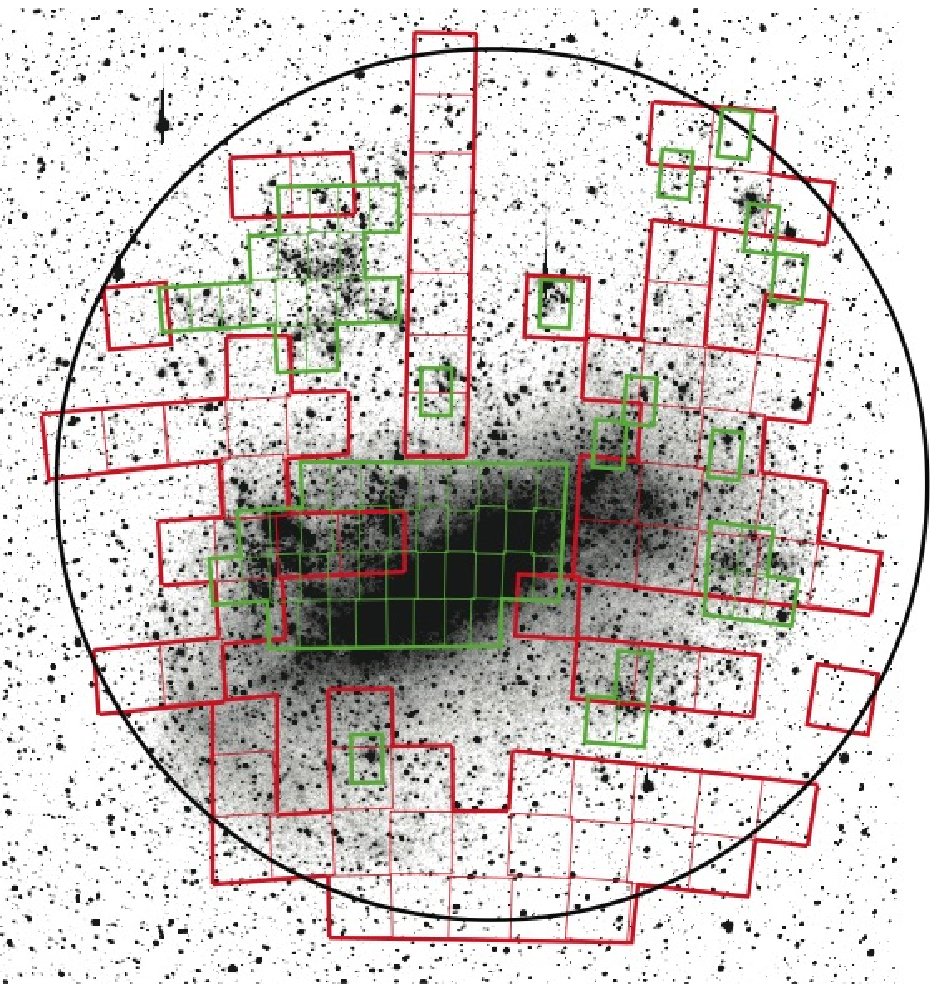}
\caption{\label{fig:LMC} Survey coverage of the LMC.  The green outlines denote the fields we observed in 2013 (Paper I), while the red  denotes the fields covered in 2014 (this paper).  The large circle shows the regions of our intended survey and has a diameter of 7$^\circ$ and is centered on $\alpha_{\rm 2000}$=5$^h$18$^m$00$^s$ $\delta_{\rm 2000}$=$-68^\circ$45\arcmin00\arcsec.The image come from the R-band ``parking lot" camera image of Bothun \& Thompson (1988).}
\end{figure}

We were assigned 10 nights for our second year of imaging, (UT) 6-10 Sep 2014, and 5-9 Nov 2014.  We were also able to obtain some observations during four additional nights during engineering tests of the new camera, on 12-13 Aug 2014 and 14-16 Aug 2014.
The seeing was unusually poor for Las Campanas during several of our nights, with the result that we repeated many fields, insisting that the seeing be adequate for the crowding in a particular field, generally $<$4.5 pixels full-width-at-half-maximum, corresponding to $<$2\arcsec, but some fields required significantly better seeing.  We also found that the image subtraction technique was badly compromised if the seeing varied by more than about 10\% on the three images, and so we repeated exposures if we were not achieving this much consistency.  Since some data were also taken through light cirrus, we also checked the photometric zero-points on each frame using the AAVSO Photometric All-Sky Survey {\it B-band} magnitudes listed in the UCAC4 catalog (Zacharias et al.\ 2013), and repeated any fields that were suspect.
In the end, we observed 137 new fields,  63 in the SMC  and 74 in the LMC, covering approximately 14.5 deg$^2$ and 17.0 deg$^2$, respectively.   The outlines of these fields are shown in Fig.~\ref{fig:SMC} and \ref{fig:LMC} in red. We also show the outlines (in green) of the regions observed in 2013 with the SITe\#3 camera.  There is considerable overlap: in some cases we wanted to double-check that we were doing as well as we did last year, and in other cases we wanted to re-observe a field that had been taken under poorer conditions.  We calculate that the total area we have so far surveyed is 17.3 deg$^2$ in the SMC and 24.8 deg$^2$ in the LMC, approximately 60\% and and 57\% of our intended survey areas in the two galaxies. 
We are scheduled for another 10 nights in 2015 November; if conditions are similar, we should readily be able to complete our survey.  

Analysis of the frames followed the same method we used in Paper I, where we employed the  image subtraction code High Order Transform of PSF And Template Subtraction ({\it \mbox{HOTPANTS}}) described by Becker et al.\  (2004). The larger field of view of the camera led to some poor results in the corners, but we have enough overlap on adjacent fields that we feel this is not a practical issue.  Along with the image subtraction, aperture photometry was run on the frames and photometrically significant candidates were identified.   The combination of the two techniques proved very valuable in identifying potential WR stars.   Accurate  (0\farcs5) coordinates were obtained by running the data through the ``astrometry.net" software (Lang et al.\ 2010) locally installed on our machines.  

Prior to attempting spectroscopic confirmation, candidates were examined through VizieR to make sure they were not already known emission-line sources; this eliminated a number of objects, mostly planetary nebulae.  (We were already suspicious of these as they had no counterparts on the continuum frames.)  In addition, inspection of the candidates' 2MASS colors allowed us to eliminate red stars, which show up as candidates as there is a strong absorption band in the continuum filter (see Paper I as well as Neugent \& Massey 2011).  In the end, we had four new candidates in the SMC, 10 new candidates in the LMC, along with three additional candidates from the first year which we had not managed to observe (all in the LMC).  One additional LMC candidate is awaiting imaging in excellent seeing as it is rather crowded.  

Spectroscopic followup was carried out on the 6.5-m Magellan Clay telescope using the Magellan Echellette (MagE) spectrograph (Marshall et al.\ 2008) during a single night of observing, (UT) 9 Jan 2015.  MagE provides full spectral coverage in the optical from 3100\AA\ to 1$\mu$m at a resolving power $R$ of 4100 using a 1\arcsec\ slit. The sky was clear, and the seeing (as reported on the guide camera) was  0\farcs7-1\farcs0.  Exposure times were typically 600-1200s, with a few longer and a few shorter.  Bias frames and various flat-field exposures (Xe-flash and dome flats) were taken, but as described in Paper I and Massey \& Hanson (2013), we found that we actually did better in terms of signal-to-noise (S/N) by {\it not} flat fielding our data, as the MagE chip is quite uniform and obtaining high signal-to-noise flats over such a large wavelength regime presents quite a challenge.  The individual orders were combined after flux calibration using observations of spectrophotometric standards.  The data were extracted using Jack Baldwin's ``mtools" IRAF routines, and wavelength calibrated and fluxed with the usual IRAF echelle reduction tasks.  

\section{New Discoveries}

Six of our candidates proved to be ``winners," with strong He~II $\lambda 4686$ emission.  Of these, four are newly found Wolf-Rayet stars, two of which belong to the newly recognized WN3/O3 type (Paper I), and two of which are ``slash" WRs (one ``hot" and one ``cool").    Of the other two stars with He~II emission, one is an odd O6.5 supergiant, and the other is an emission-line star with very peculiar line profiles.  Among the ``losers" (stars without He~II $\lambda 4686$ emission), four are late-type stars which we failed to eliminate before observing, and which will not be discussed further here.   Six are previously unknown B-type stars.  These stars lacked emission at He~II~$\lambda 4686$, and were detected at marginal levels; we included these in our spectroscopy followup in order to make sure we were not missing any weak-lined WRs.  One other candidate had been identified as a RR Lyr variable that was detected at a high significance level.  Although variability could produce a false WR candidate (if the star was fainter in the continuum exposure), we observed this to be sure it had been correctly identified as an RR~Lyr star.  The spectrum was that of an A-type star, as would be expected for an RR~Lyr.  

\subsection{New Wolf-Rayet Stars}

Four of our LMC candidates proved to be Wolf-Rayet stars.  We list their identifications and properties in Table~\ref{tab:WRs}.  Two of these stars are additional examples of the WN3/O3 class that we introduced in Paper I, bringing the total of this type to eight.  The other two are ``slash" stars  (see, e.g., Bohannan \& Walborn 1989, Crowther \& Smith 1997, and Crowther \& Walborn 2011). In general, we rely upon the expanded classification system of Smith (1968) as summarized in Table 2 of van der Hucht (2001), as this provides consistency not only with the Milky Way WRs (van der Hucht 2001), but also with most previous classifications of LMC and SMC WRs (BAT99 and Massey et al.\ 2003), as well as recent studies of the M33 and M31 WR populations (Neugent \& Massey 2011 and Neugent et al.\ 2012b, respectively).    The total number of  WRs known in the LMC is now 152, of which 13 (8.6\%) were found as a result of our survey\footnote{This number includes the 134 WRs cataloged by BAT99, minus the two demotions and 7 additions discussed by
Neugent et al.\ (2012a).    Hainich et al.\ (2014) argues that two of these additions are uncertain, namely Sk$-69^\circ$194 (B0~Ia+WN, according to Massey et al.\ 2000), and  LH90$\beta$-6 (B~I+WN, also according to Massey et al.\ 2000).  We plan to resolve this controversy in the near future.  The number also includes the nine newly found WRs in Paper I and the four discussed here, but does not include HD~38489, which we speculate in Paper~I could be a B[e]+WN binary.}.  We discuss the four newly found WRs here.

\begin{figure}
\epsscale{0.48}
\plotone{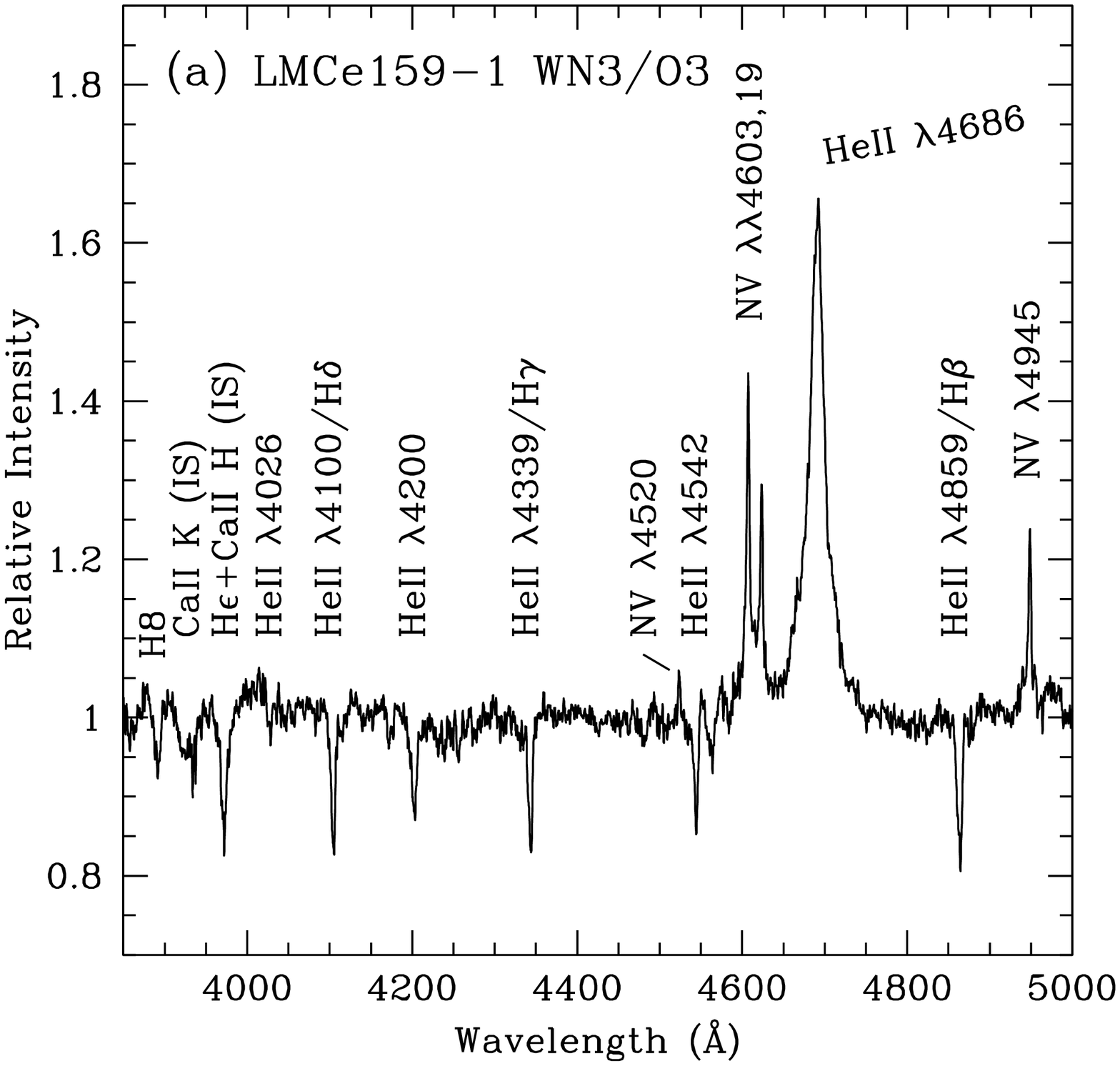}
\plotone{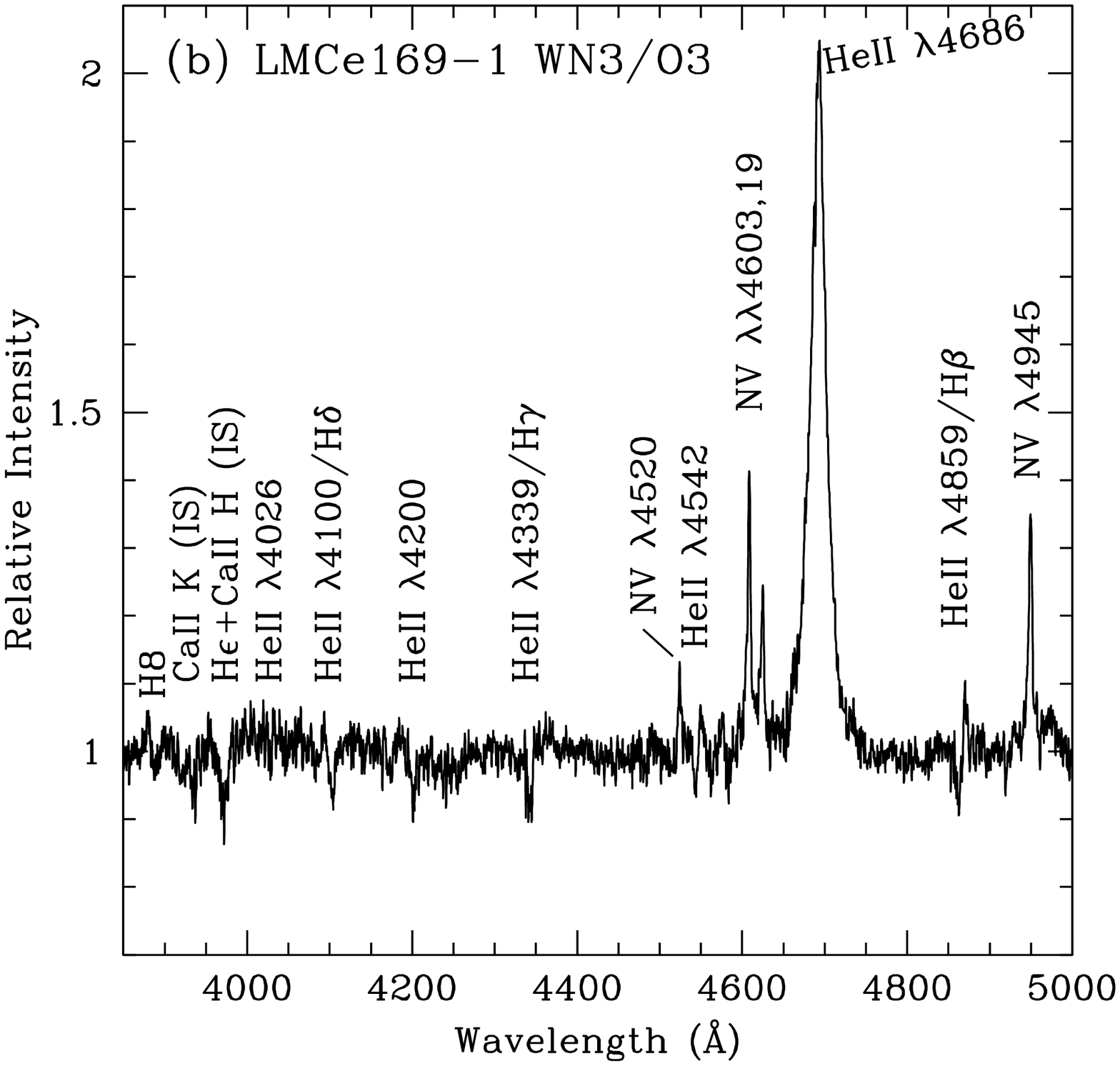}
\plotone{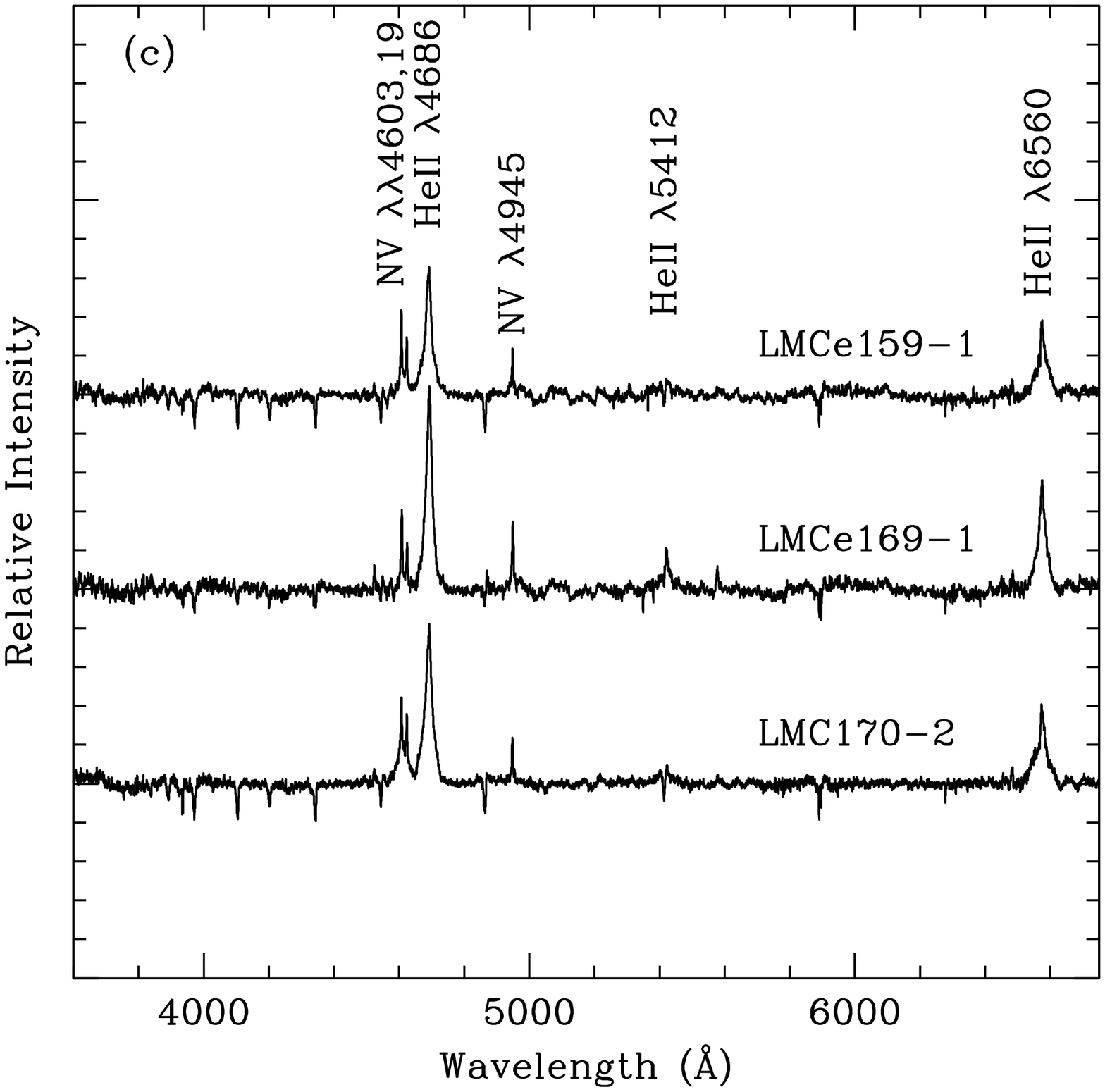}
\caption{\label{fig:O3s} Normalized spectra of two newly found WN3/O3 WRs. 	The blue portions of our spectra are shown in the top two figures for (a) LMCe159-1 and (b) LMCe169-1.  Note that the absorption is weaker, and the emission stronger, in LMCe169-1.  In (c) we compare the spectra of these two newly found WN3/O3s with each other and with the WN3/O3 star LMC170-2 discussed in Paper I and Massey et al.\ (2015).}
\end{figure}

\paragraph{LMCe159-1 and LMCe169-1, two WN3/O3s stars.} Two of our newly found WR stars are of the type we have begun calling WN3/O3, as described in Paper~I.  We show their spectra in Fig.~\ref{fig:O3s}, and their identification and properties in Table~\ref{tab:WRs}.   Like the six WN3/O3 stars we discovered in Paper I, LMCe159-1 and LMCe169-1 show H and He~II absorption but no He~I.  N~V is strongly in emission (Figs.~\ref{fig:O3s}a and b).   Note that the emission is stronger and the absorption a bit weaker in LMC169-1.  In Fig.~\ref{fig:O3s}(c) we compare the spectra of our two new discoveries to the spectrum of LMC170-2, one of the prototypes of the WN3/O3 class that we discuss in Paper I and Massey et al.\ (2015).  The presence of hydrogen in the absorption spectrum is apparent by comparing the strengths of the odd-n Pickering He~II lines (He~II $\lambda \lambda 4026, 4200, 4542$) with the even-n lines, which are coincident with the Balmer hydrogen lines.  (Our modeling of LMC170-2 suggests a He/H number ratio of $\sim$1.0.)

One of the chief arguments against the WN3/O3 stars in Paper I being WN3+O3~V binaries was that these stars were about as faint as we would expect for an early-type WN star (such as a WN3), but that they were much too faint to also include an O3~V star.   As we show in Table~\ref{tab:WRs}, the same is true here.  These new additions are even fainter than the stars in Paper I, with absolute visual magnitudes $M_V$ corresponding to $-2.6$ and $-1.8.$  These can be compared to the average absolute
magnitudes of early-type LMC WNs given by Hainich (2014):  $-3.8$ for WN3s and $-2.7$ for WN2s.  In fact, LMCe169-1 may be the faintest WR star known in the LMC (or anywhere); the only other contenders are BAT99-23, a WN3 star described as having hydrogen in its spectrum, and BAT99-69, a WC star in a crowded region whose identification may be confused; we are indebted to Brian Skiff for his notes on the latter. 
In calculating the $M_V$ values we have assumed typical reddening for LMC OB stars (see Massey et al.\ 1995), i.e., $A_V=0.4$ based upon an average $E(B-V)$ of 0.13.  This is quite consistent with the observed $B-V$ colors of these stars based upon the photometric survey of Zaritsky et al.\ (2004). These stars are about 3~mags too faint to include an O3~V component, which typically has $M_V$$\sim$ $-5.4$ (Conti 1988).  We comment further on this in Section~\ref{Sec-End}.

\begin{figure}
\epsscale{0.48}
\plotone{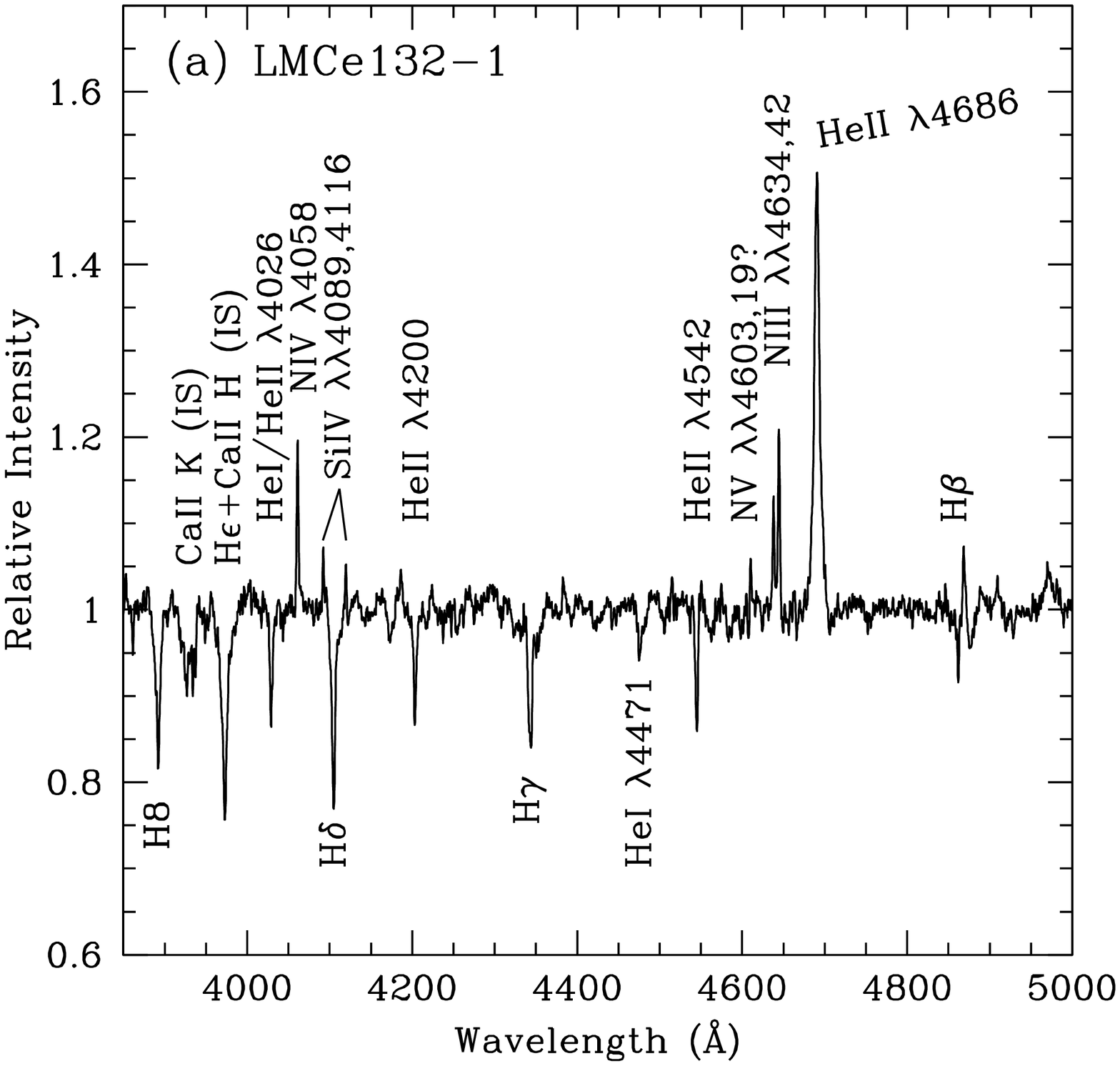}
\plotone{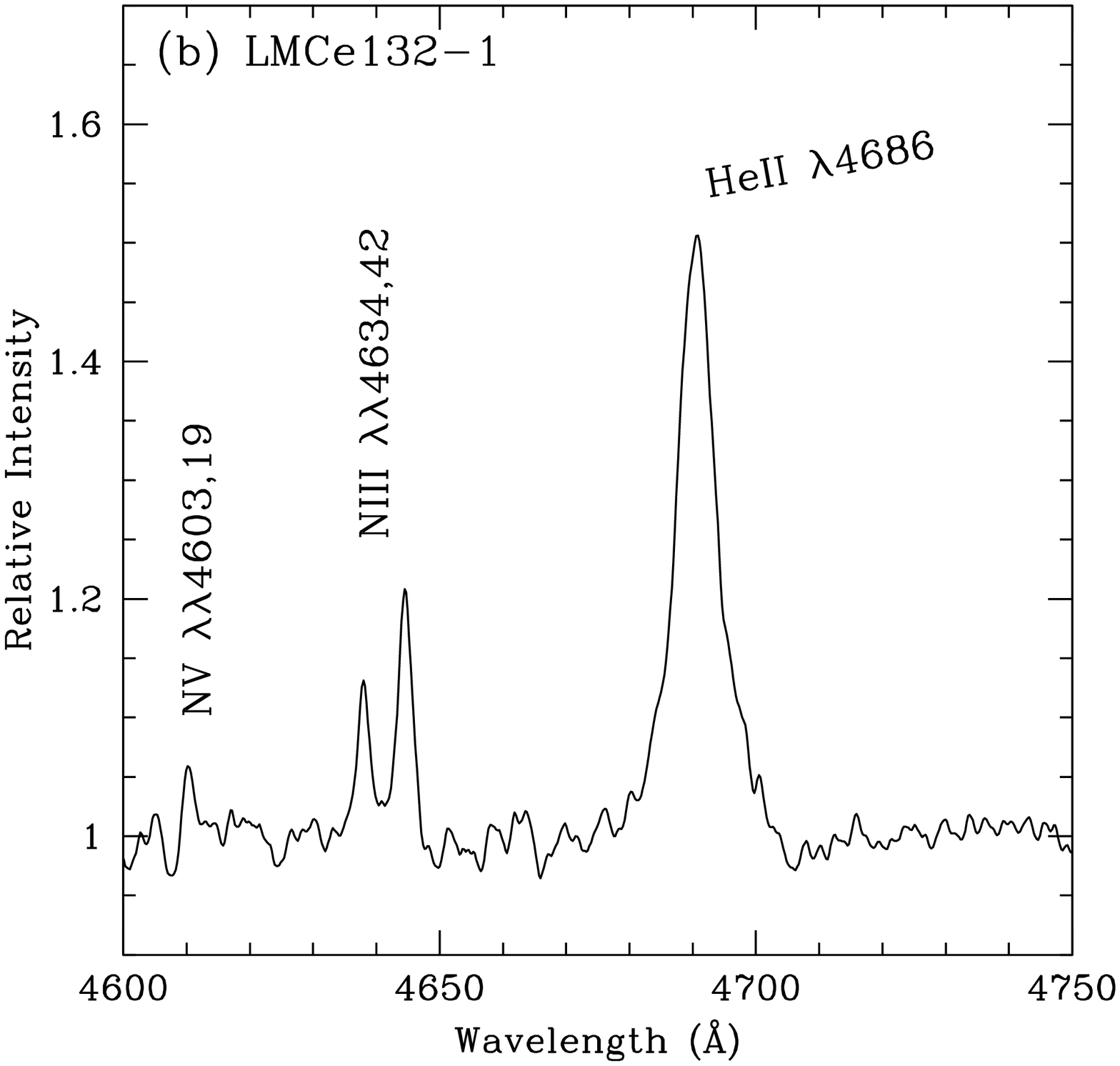}
\plotone{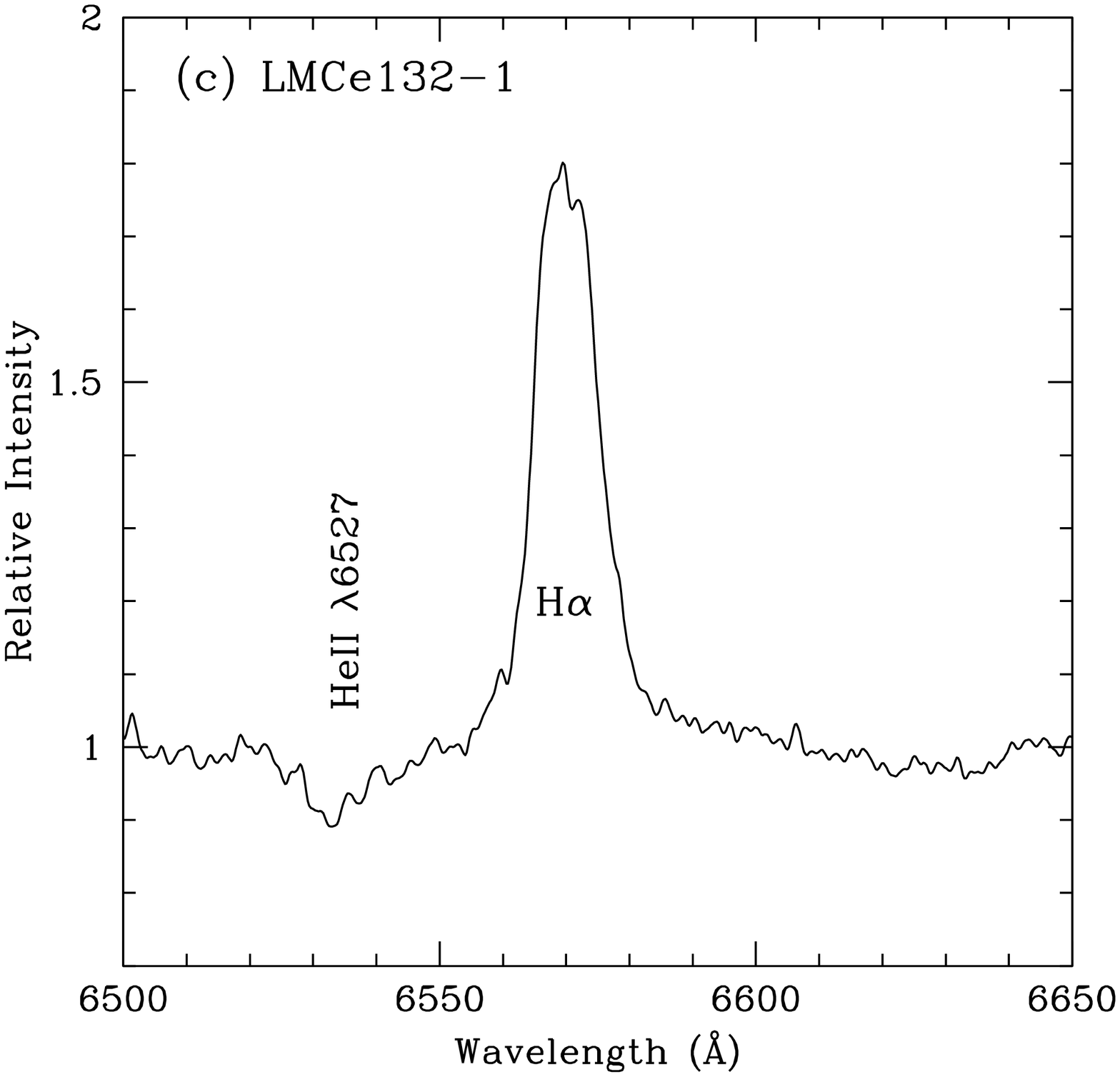}
\caption{\label{fig:LMCe132-1} Normalized spectrum of LMCe132-1, a newly found O3.5 I*/WN7 ``slash" WR. In (a) we show the region from 3850\AA\ to 5000\AA\ with the principal lines indicated.  
Note in particular the P Cyg profile in H$\beta$ which results in this star being called a hot ``slash" WR.  The broad absorption feature just to the red of the H$\beta$ emission is the diffuse interstellar band at 4881\AA. In (b) we expand the region around N~III $\lambda 4634,42$ and He II $\lambda 4686$.   In (c) We show the region around H$\alpha$.}
\end{figure}

\paragraph{LMCe132-1, an O3.5~If*/WN7 star.} The blue classification region of the spectrum is
shown in Fig.~\ref{fig:LMCe132-1}a.   The hybrid O3~If*/WN6 was introduced by Walborn (1982) to describe the spectrum of Sk$-67^\circ$22, which became the prototype for the ``hot" slash stars that show spectral characteristics intermediate between O3~If stars and WN6 stars. 
 Crowther \& Walborn (2011) subsequently refined this, using the presence of a P Cygni profile in the H$\beta$ line to require this intermediate classification, just as we find here.  Using their criteria, we classify this star as O3.5~If*/WN7.  True to the classification criteria given in their Table 2, there is even extremely weak NV $\lambda 4603,19$ emission present 
 (Fig.~\ref{fig:LMCe132-1}b).  
H$\alpha$ is strongly in emission, as shown in Fig.~\ref{fig:LMCe132-1}(c).
The photometry is consistent with normal reddening.
 
 Such hot ``slash stars" are likely hydrogen-burning objects.  Should they even be counted as WRs?   We note that the BAT99 contains nine examples of such objects, including Sk$-67^\circ$22 (BAT99-12), the prototype of this class.  None of these have been ``demoted" to Of-type status, although a fresh look at these stars in the light of the Crowther \& Walborn (2011) study is probably warranted.  If these nine stars count as WRs, so should LMCe132-1.
 
 \paragraph{LMCe063-1=Sk$-$69$^\circ$240, a WN11 star.}  
 This star was recognized as an emission-lined object by Henize (1956), who cataloged it as LHA 120-S131.  It was classified as a Be star as part of the objective prism survey of Sanduleak (1970), where it is listed as Sk$-69^\circ240$, but it subsequently became clear that whatever the star is, it is not a  Be star.   Shore \& Sanduleak (1984) obtained a better optical spectrum, as well as an ultraviolet  spectrum with {\it IUE}.  They describe the optical spectrum as a pure emission-line spectrum (without P Cygni profiles), with the Balmer lines in emission to $n=10$, He~I emission, and He~II $\lambda 4686$ and N~III~$\lambda \lambda 4634,42$ in emission.     The {\it IUE} spectrum showed the usual
stellar wind lines (Si IV $\lambda 1400$, C~IV~$\lambda 1550$, N~V~$\lambda 1245$, and He~II~$\lambda 1640$).  A high dispersion ($R$$\sim$7000) photographic optical spectrum is shown in Fig.~28 of Stahl et al.\ (1985), who describe the line profiles as ``unique," with the H and He~I emission split by a sharp central absorption. 

\begin{figure}
\epsscale{0.48}
\plotone{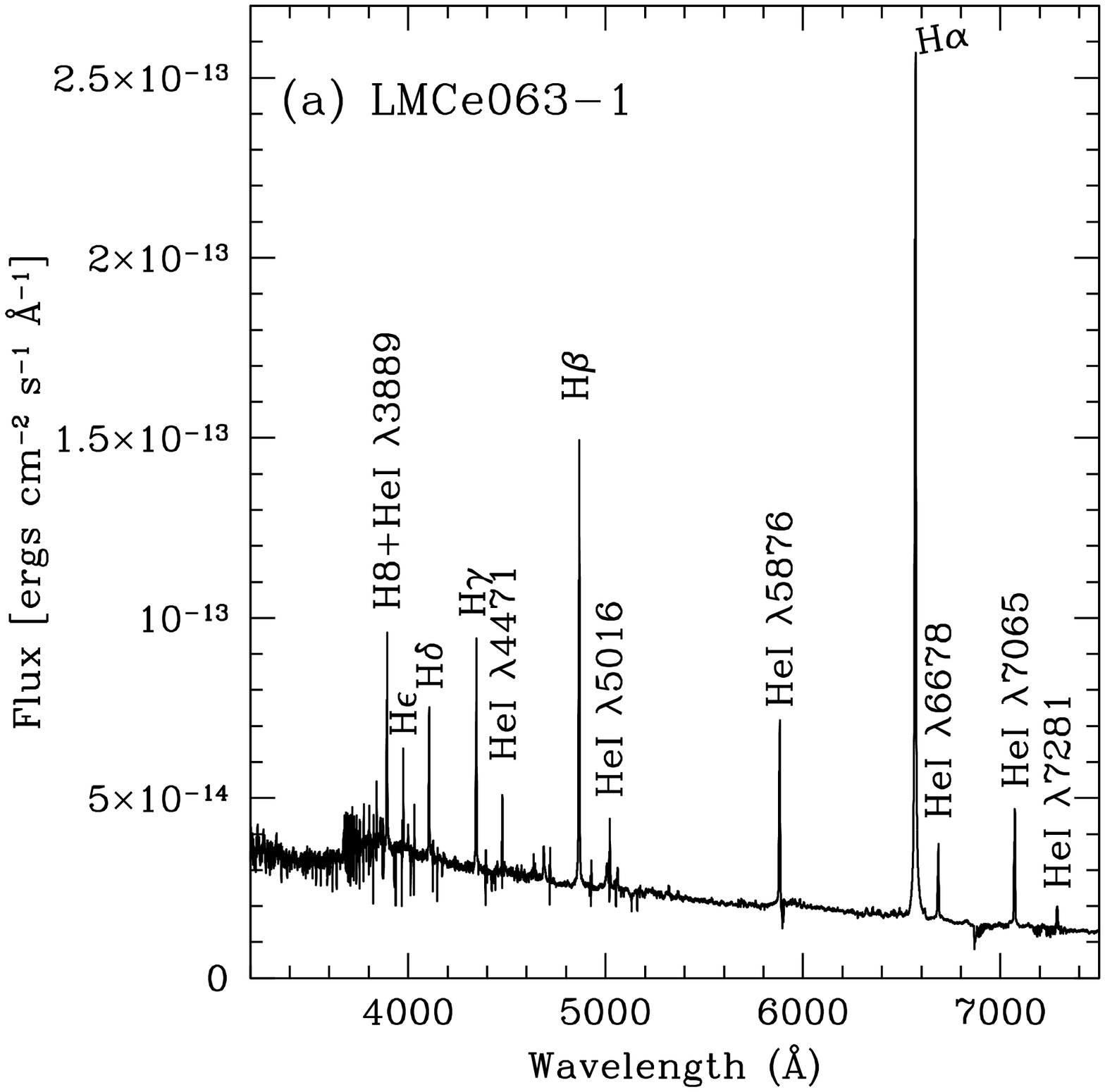}
\plotone{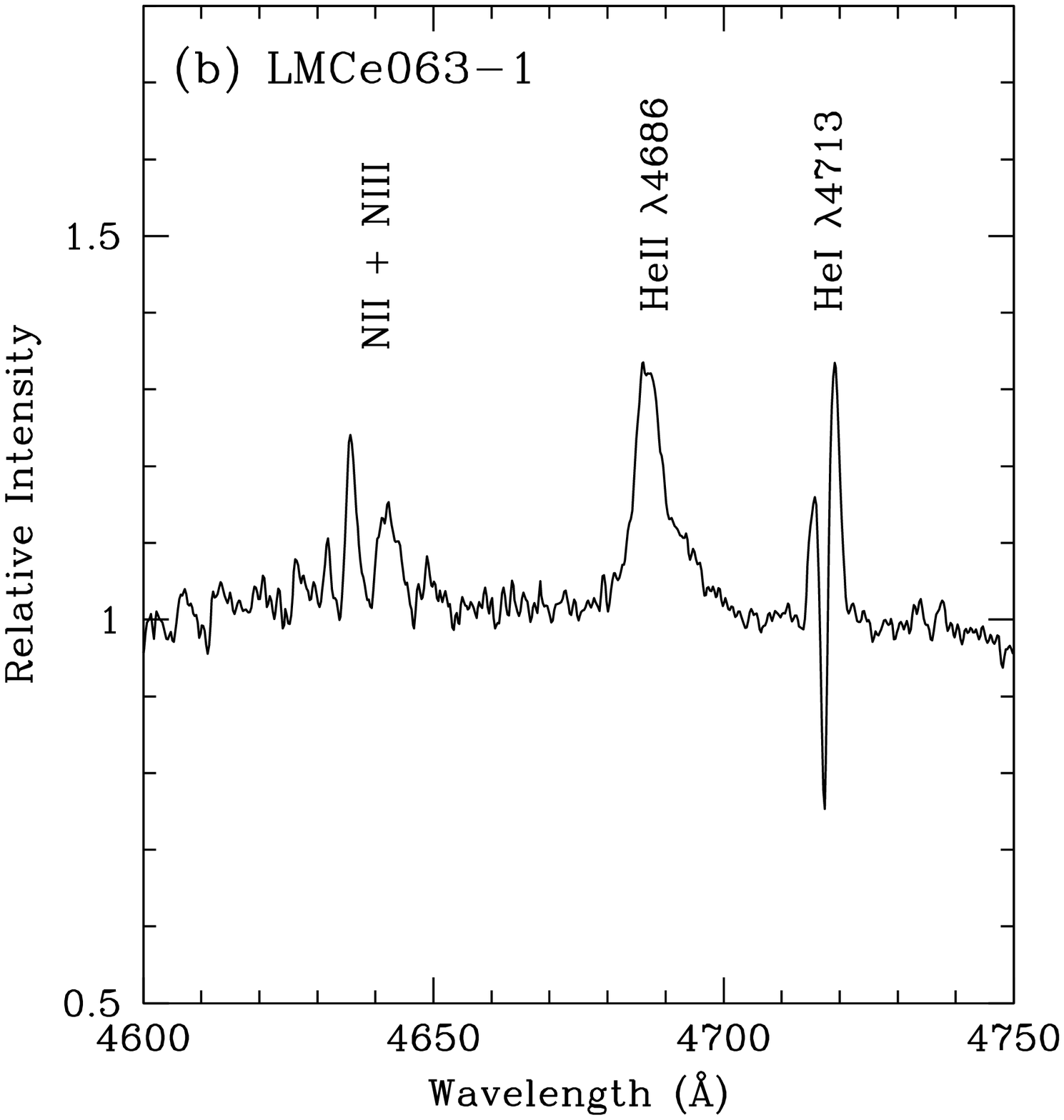}
\plotone{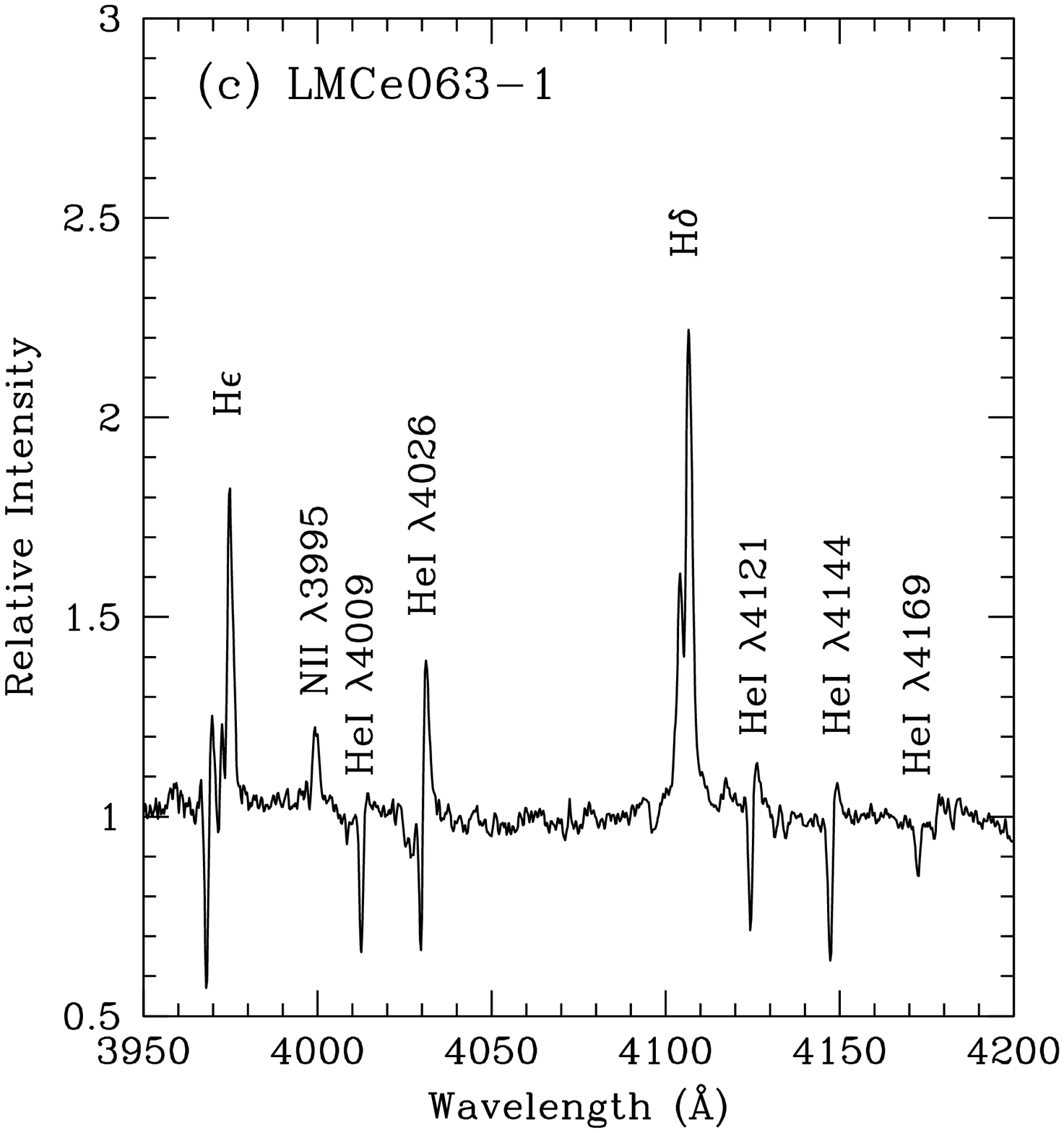}
\plotone{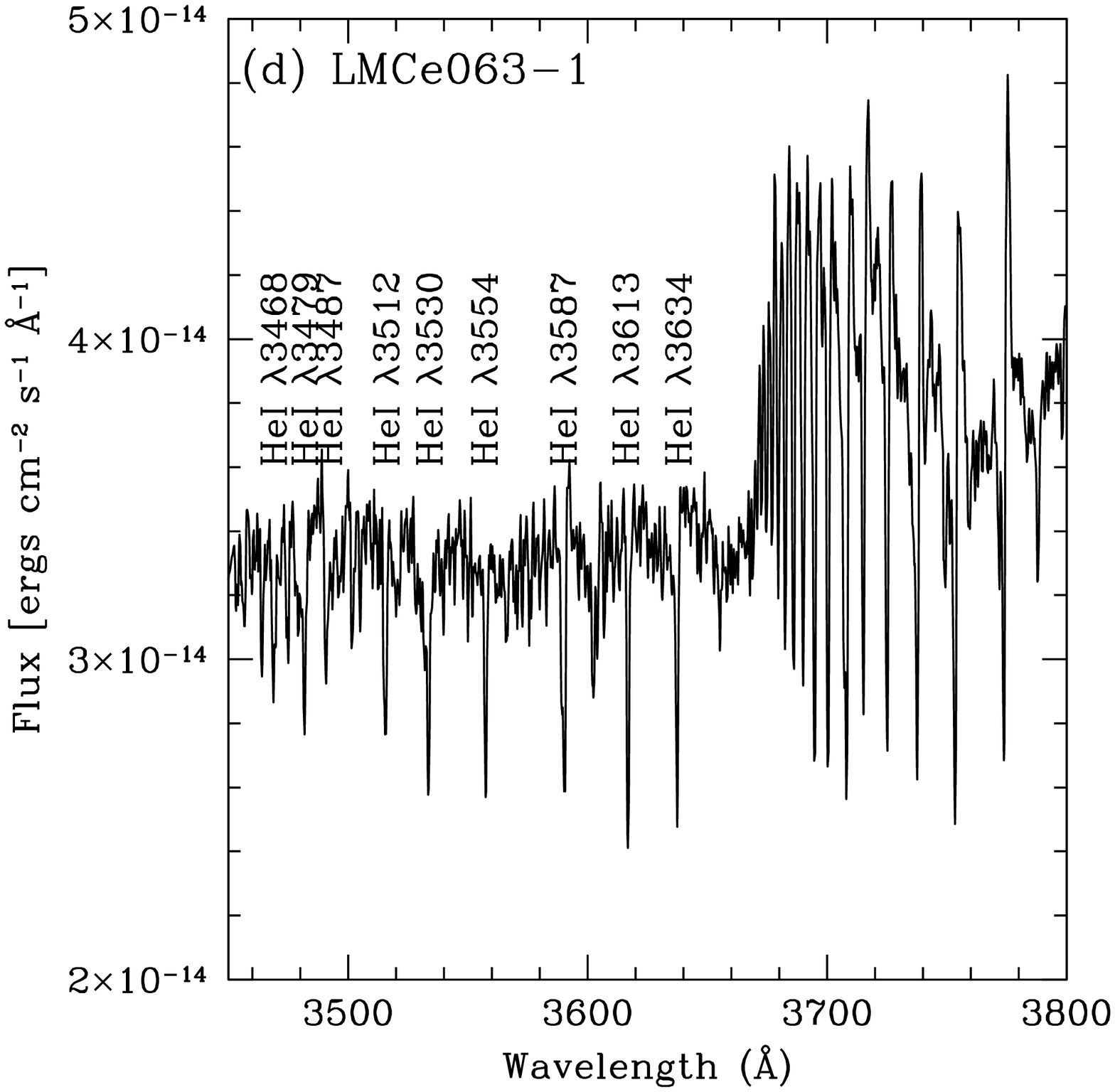}
\caption{\label{fig:LMCe063-1} Fluxed and normalized spectra of LMCe063-1, a WN11 star.  In (a) we show an overview of the spectrum from 3200\AA\ to 7500\AA, with the principal lines identified.  In (b) we show the region around He II $\lambda 4686$. Note the example of the one of the ``unique profiles" seen in He I $\lambda 4713$; this is typical of many of the He I lines longwards of the Balmer jump, plus the lower Balmer lines.  (d)  Shortwards of the Balmer jump we find a series of pure absorption He~I lines; note also the strong P Cygni profiles in the upper Balmer lines.}\end{figure}

 Our own spectrum of this interesting object is shown in Fig.~\ref{fig:LMCe063-1}. Strong He I and hydrogen lines dominate the spectrum (Fig.~\ref{fig:LMCe063-1}a).  He~II$~\lambda 4686$ emission is indeed weakly present, but we were suspicious of the identification of the neighboring feature as N~III~$\lambda 4634,42$, given the very choppy appearance (Fig.~\ref{fig:LMCe063-1}b).    Instead, we realized this emission is dominated by N~II, as witness the strong N~II $\lambda 3995$ line we
 find in the blue (Fig.~\ref{fig:LMCe063-1}c).  
 
 Comparing this star to others in the literature, we were struck by the similarity to He3-519 (Walborn \& Fitzpatrick 2000) and HDE 269582 (Bohannan \& Walborn 1989).  Both of these are examples of ``cool" slash stars, what were originally called Ofpe/WN9 stars, but are generally classified as WN9-11 today (see, e.g., Crowther \& Smith  1997).  Following this, we would classify the star as WN11.   We note that the WN11 stars have a strong linkage to luminous blue variables (LBVs) (Crowther 1997);  indeed,  Walborn et al.\ (2012) reported that HDE~269582 is currently undergoing a major LBV outburst. The one way our spectrum differs significantly from ``normal" WN11s is the presence of the peculiar absorption lines splitting the hydrogen and strong He I emission lines, perhaps indicating the presence of a disk. (There are some similarities to the Oe stars discussed by Sota et al.\ 2014; we are grateful to Nolan Walborn for pointing this out.)
 
Our wavelength coverage extends further into the blue and at higher signal-to-noise than is usually the case for such stars, and we'll call attention that there are {\it some} pure absorption lines in the spectrum, as shown in Fig.~\ref{fig:LMCe063-1}d.  We see here a sequence of He I lines, mostly triplets from the $1s2p$ $^3$P$^o$ - 1s$n$d $^3$D, with $n\geq8$ ($n$=8 corresponding to He I $\lambda 3634$ line).  It is interesting that these high-order lines are in absorption, while lower members occurring at wavelengths above the Balmer jump are not (e.g., He~I~$\lambda 4471$, which is the $1s2p$ $^3$P$^o - 1s4d$ $^3$D transition). We also find in this figure a sea of P Cygni profiles to the upper Balmer lines; we can count upper Balmer lines to H28.

\subsection{Other Emission-Line Stars}

Besides the four WRs discussed above, our spectroscopy has identified two additional emission-line stars in the LMC.   We list the properties of these stars in Table~\ref{tab:others}.  In each case, a simple classification eludes us.  We discuss each of these stars below.

\paragraph{LMC156-2, an Onfp star.}

\begin{figure}
\epsscale{0.48}
\plotone{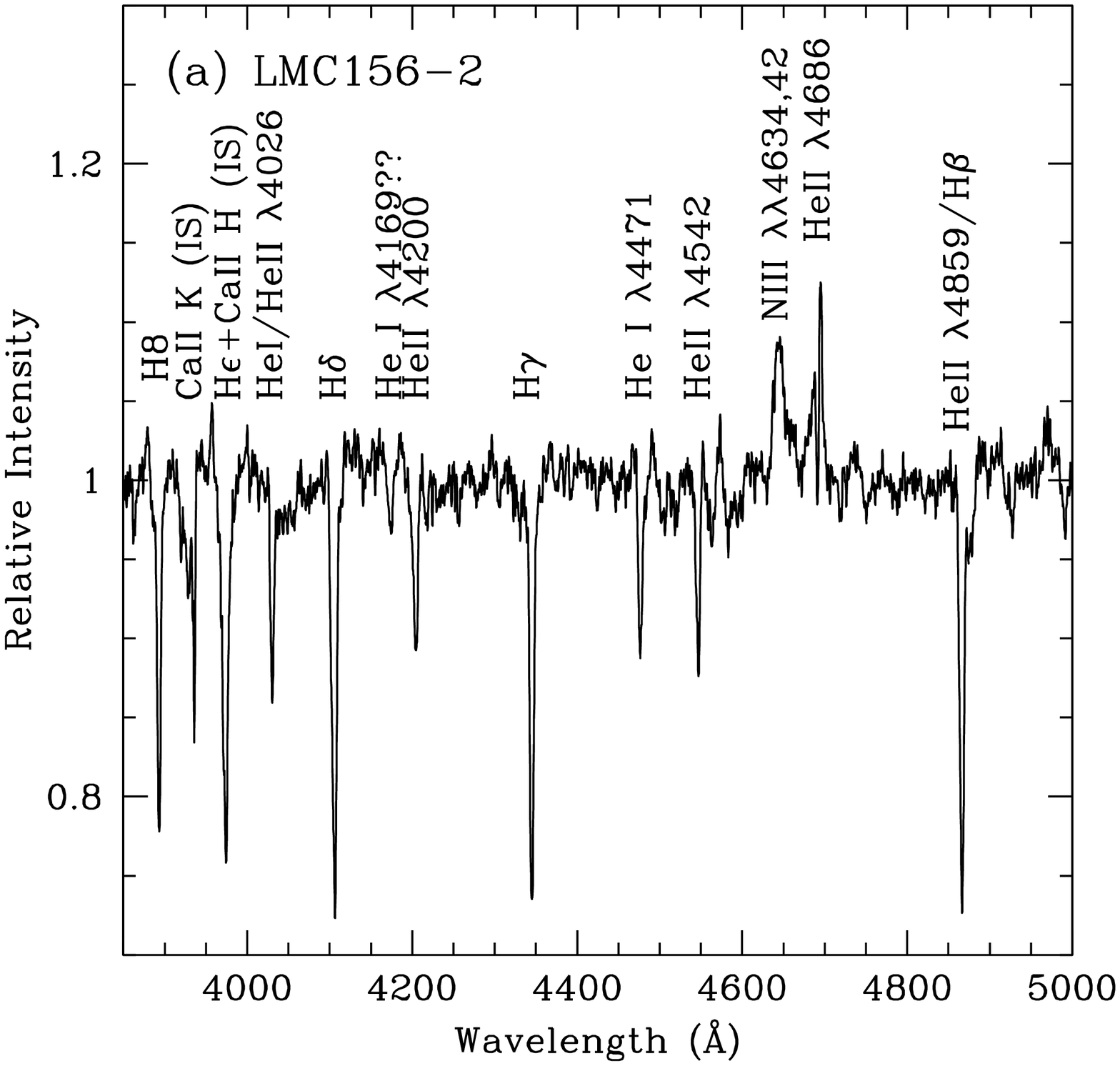}
\plotone{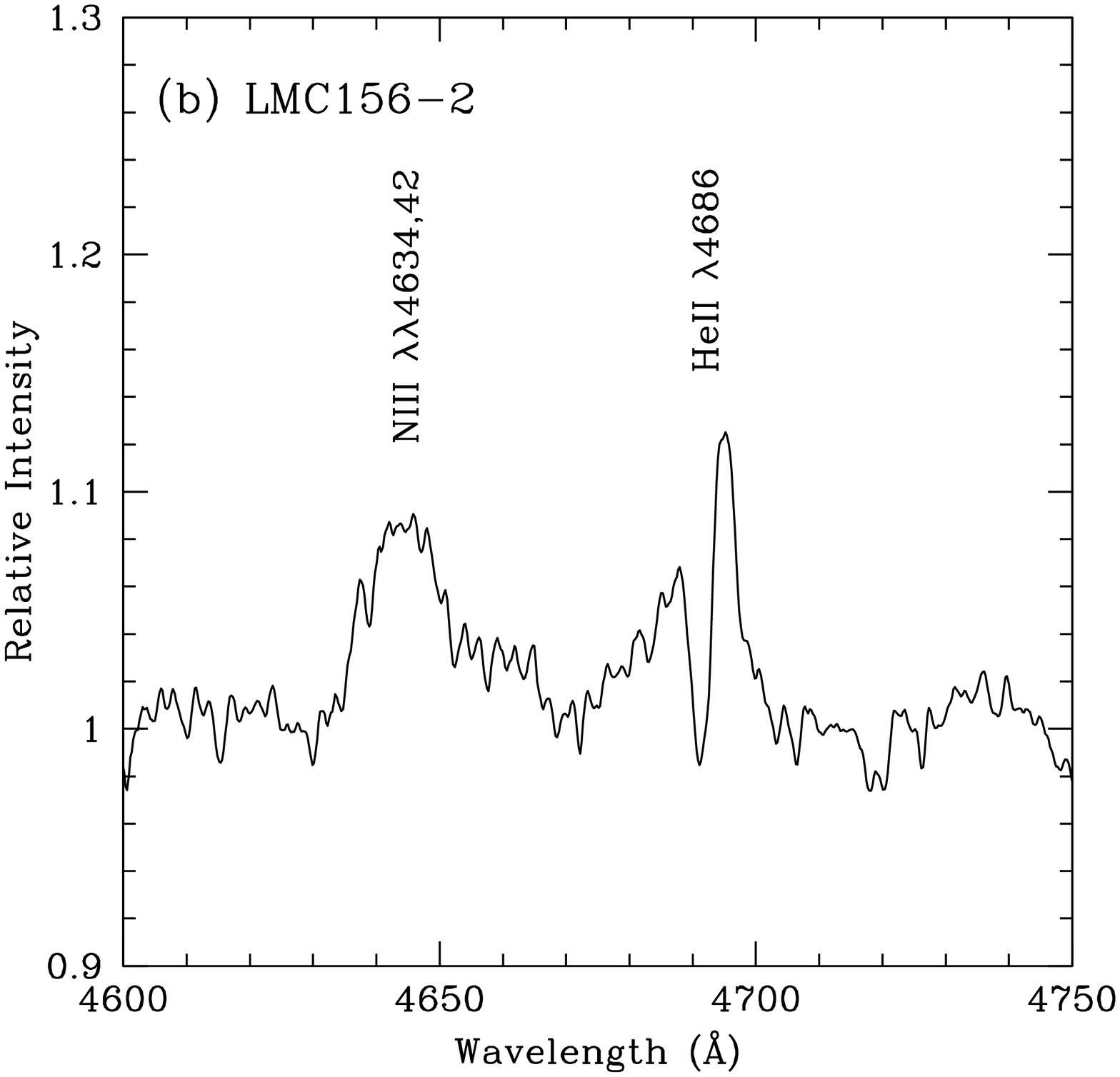}
\plotone{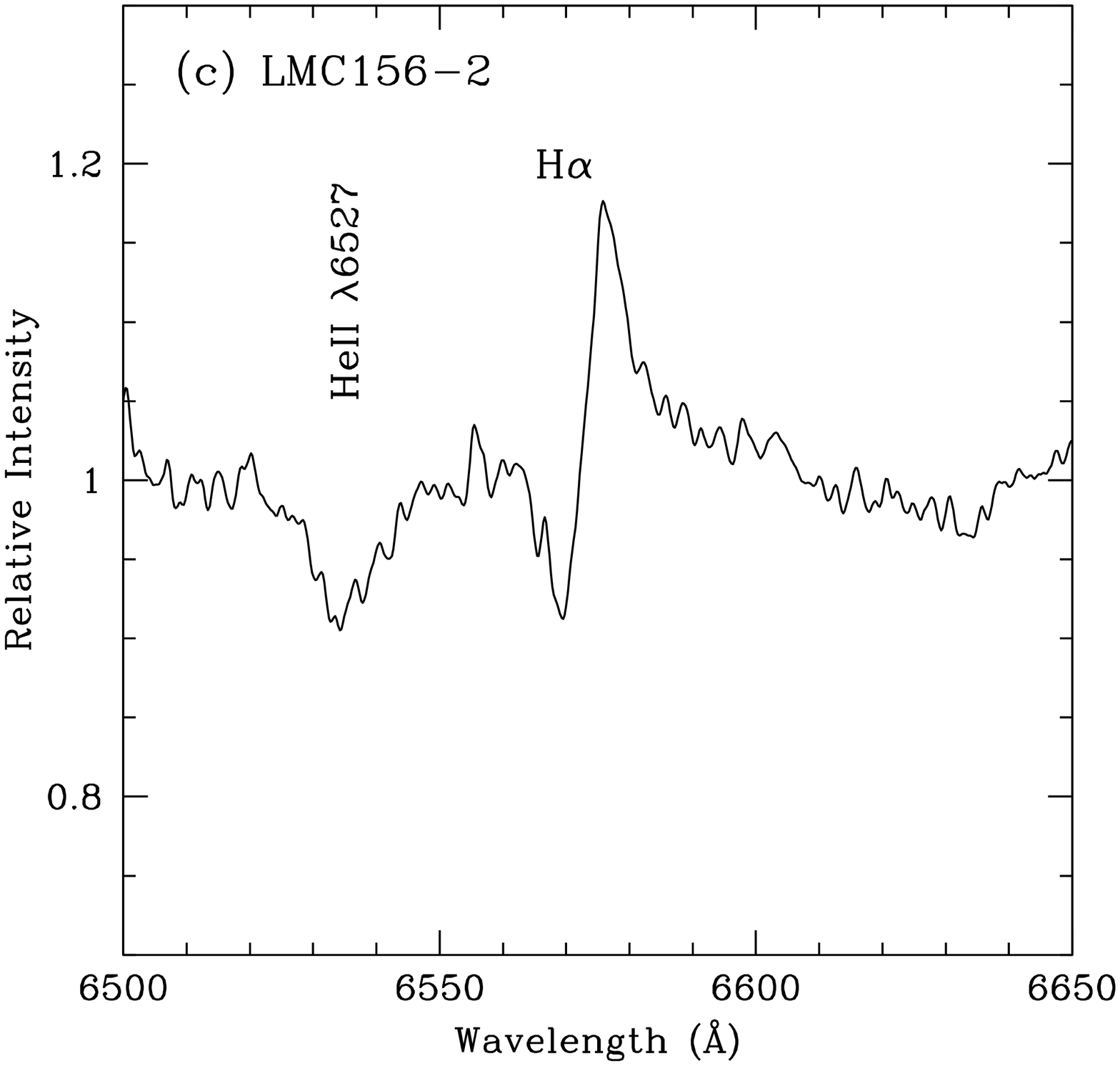}
\caption{\label{fig:LMC156-2} Normalized spectrum of LMC156-2, an Onfp star.  In (a) we show the region from 3850\AA\ to 5000\AA\ with the principal lines identified.  In (b) we expand the region around N~III $\lambda 4634,42$ and He II $\lambda 4686$.   In (c) We show the region around H$\alpha$.}
\end{figure}

The spectrum of this star is shown in Fig.~\ref{fig:LMC156-2}.  At first blush, the star seems to be an O6.5 If, with He~I~$\lambda 4471$ a bit weaker than He~II~$\lambda 4542$, and the supergiant ``If" luminosity classification due to strong N~III~$\lambda \lambda 4634,42$ and He~II $\lambda 4686$ emission.   However, a closer examination reveals multiple strange things: as shown in Fig.~\ref{fig:LMC156-2}(b), N~III~$\lambda \lambda 4634,42$ is anomalously broad, and the He~II~$\lambda 4686$ emission shows a strong absorption component on the bluewards side, a classic P~Cygni profile.  The He~II $\lambda 4686$ line is formed in the stellar winds in O-type supergiants, and is a good mass-loss indicator, as is H$\alpha$, and we see in Fig.~\ref{fig:LMC156-2}(c) that the H$\alpha$ profile is similarly afflicted. Such stars were first classified as ``Onfp"  by Walborn (1973), and Magellanic Cloud members of this class have been discussed more recently by Walborn et al.\ (2010).  The exact nature of these stars is not well understood; Walborn et al.\ (2010) finds that most of members of this group are binaries. Measurements of the radial velocities would tend to support this in the case of LMC156-2 as well: the He~II $\lambda \lambda 4200, 4542$ lines have a radial velocity about 70 km s$^{-1}$ more negative than that of He~I~$\lambda 4471$, with the Balmer lines of intermediate velocity.   

\begin{figure}
\epsscale{0.48}
\plotone{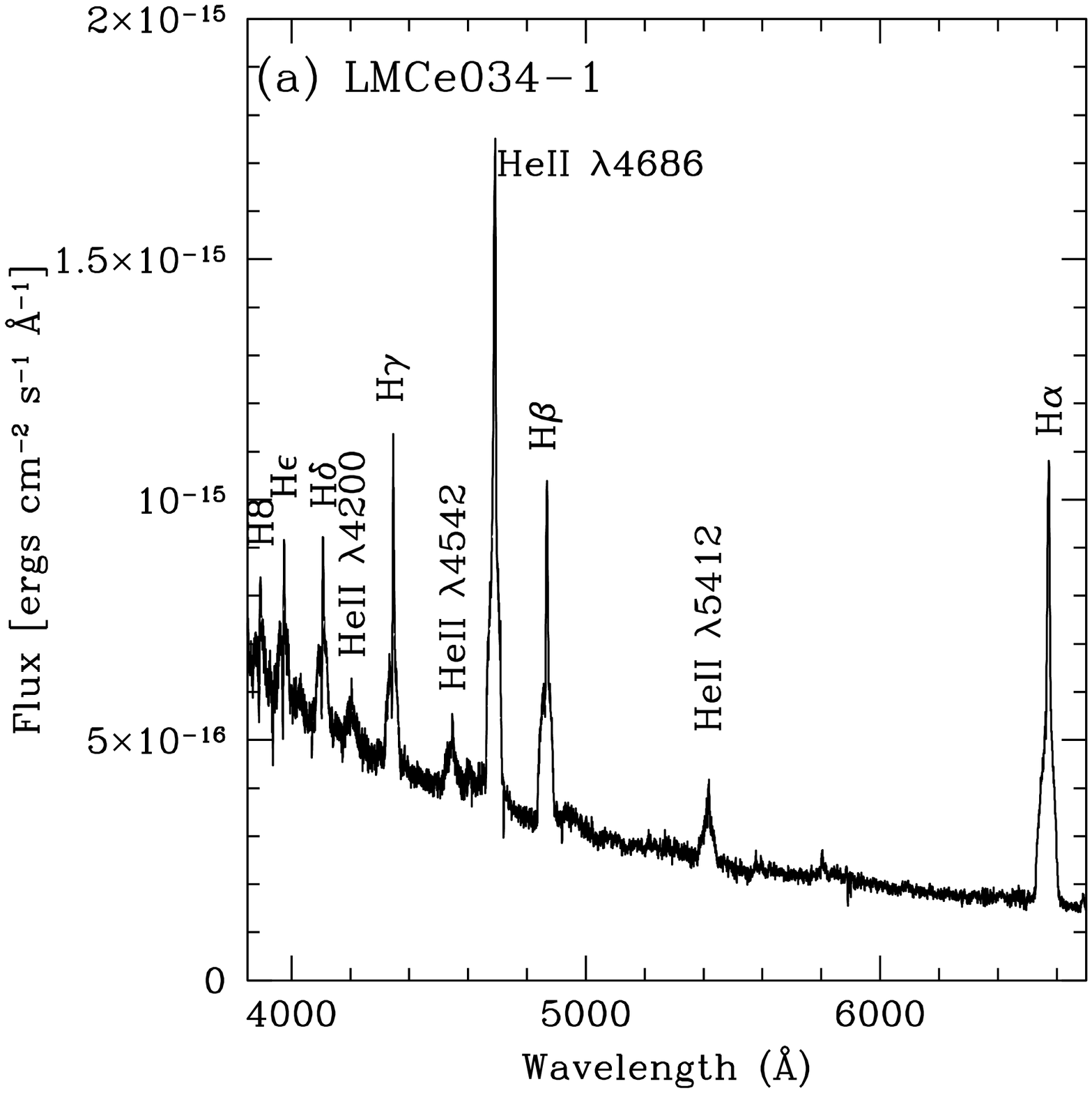}
\plotone{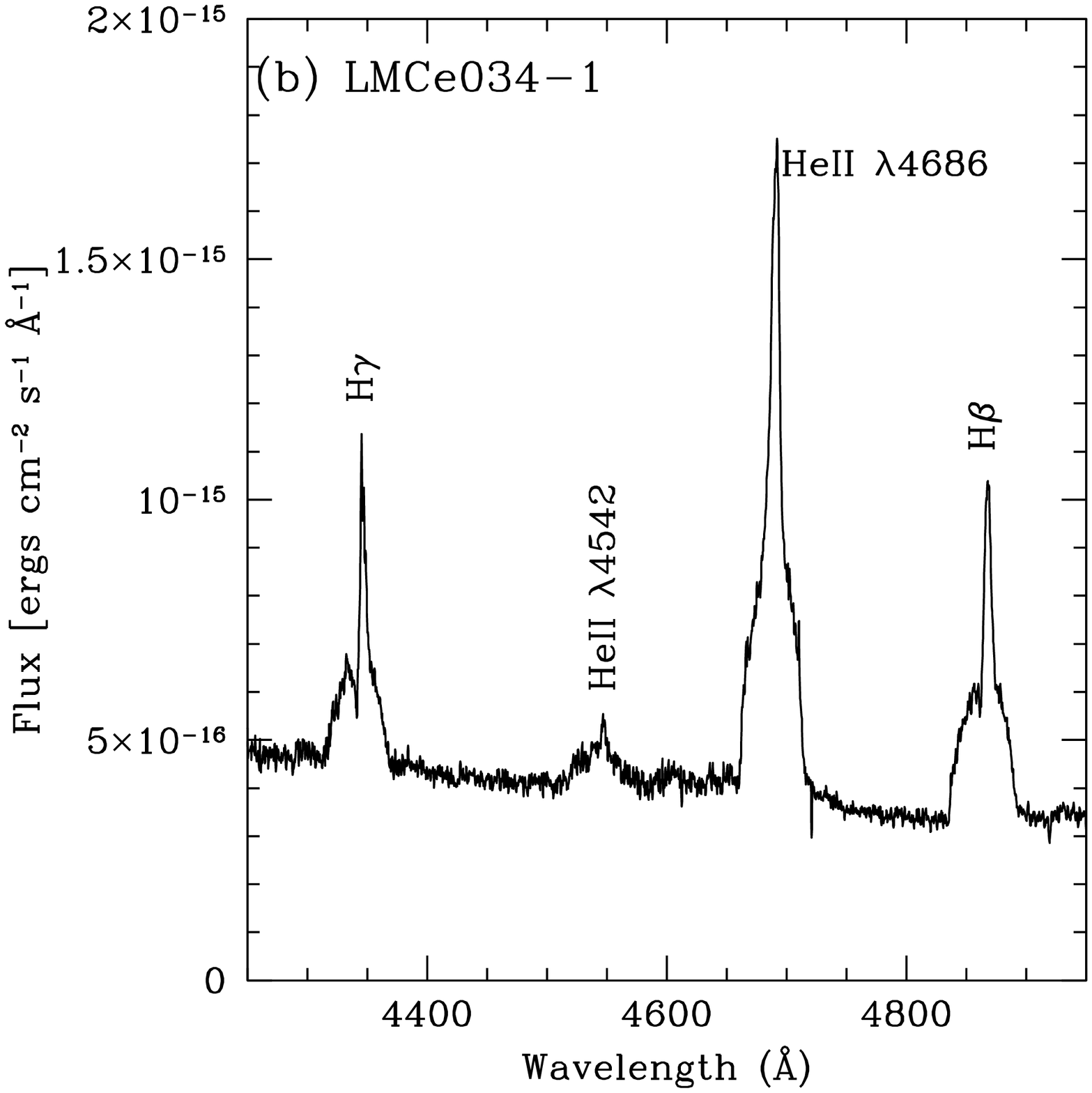}
\caption{\label{fig:LMCe034-1} Fluxed spectrum of LMCe034-1, a peculiar emission-line star.
In (a) we show an overview of the spectrum from 3850\AA\ to 6700\AA, with the principal lines identified.  In (b) we expand a region to show the peculiar line profiles.}
\end{figure}

\paragraph{LMCe034-1 (unknown type emission-line star).}  The most interesting find besides our new WRs has been that of this very peculiar emission-line star.  We give an overview of its spectral features in Fig.~\ref{fig:LMCe034-1}(a).  The spectrum is dominated by hydrogen and He~II emission lines; indeed, the strongest line is that of He~II~$\lambda 4686$.  Fig.~\ref{fig:LMCe034-1}(b) shows that the profiles consist of a skinnier component on top of  a very broad component.  The base of the He~II~$\lambda 4686$ line extends from 4660\AA\ to 4716\AA, or $\pm1800$ km s$^{-1}$!  The narrower component is only skinnier by comparison,  extending $\pm550$ km s$^{-1}$.  Note the sharp absorption features on the Balmer lines, getting progressively stronger towards higher lines.  These occur on the blue edge of the narrower emission, and are somewhat reminiscent of the ``unique" profiles we discussed in regards to LMCe063-1.   We do not see any evidence of  N~V $\lambda 4603,19$, N~IV $\lambda 4058$, N~III $\lambda 4634,42$, or the N~II complex 4600-4650\AA, despite the good signal-to-noise (50 per spectral resolution element) in these regions.

LMCe034-1 is a member of a 2\farcs1 pair, and because of this, photometry in the literature is a little ambiguous.  In Table~\ref{tab:others} we have used the ``continuum" magnitude that we measure from our spectrophotometry, converted to the Vega (standard {\it UBVRI}) system. 
We do find that this star varies in light.  We took two exposures of this field, once on (UT) 2014 Sept 10  under poor conditions, and once on (UT) 2014 Nov 11.  We measured LMCe034-1 as $0.90\pm0.03$~mag fainter on the continuum image in the first exposure as in the second.  It is also possible that the He~II~$\lambda 4686$ emission was weaker based on the {\it WN} image, but that is harder to judge, given the poorer quality of the data.

What do we make of this peculiar object?  We can rule out it being an AM~Her cataclysmic variable based on its radial velocity and apparent brightness.   John Hillier, Howard Bond, and Nolan Walborn were all kind enough to comment on the spectrum.  Although broad components like this are often attributed to Thompson scattering, in this case the profiles look too boxy for that explanation.  Instead, Hillier has offered the useful conjecture that we may be looking at a two-component wind.  After all, the skinner component is none too skinny.  A binary explanation can not be ruled out, and we will continue to monitor this star for radial velocity variations.  (We saw none in two spectra separated by 5.5 hours.) Nevertheless, the fact that both the broad and narrow components show up in most of the lines suggests that this is an unlikely explanation for the peculiar spectrum, regardless of whether the star is a binary (with an unseen companion) or not.  Walborn has
instead noted the similarity to NGC 1624-2 (Wade et al.\ 2012), and has suggested that the strange profiles
are due to a combination of photospheric and magetospheric components.

\subsection{Non-Emission Line Stars}

As mentioned above, six of our candidates turned out to be normal B-type stars.   These candidates were marginal, having magnitude differences between the WN and CT filters of $\sim$0.1~mag or less. Given the number of stars involved, our methodology is bound to produce a few false positives. We give the identifications of these B stars in Table~\ref{tab:losers} for completeness, but do not discuss these in detail.  We were guided by the Walborn \& Fitzpatrick (1990) atlas in the classification.  

LMCe019-1 was identified as an RR Lyr variable by OGLE (LMC-RRLYR-06881) with a maximum magnitude of $I=15.72$, an amplitude of 0.8~mag, and a period of 0.5 days (Soszy\'{n}ski et al.\ 2009).  Our spectrum shows this to be of A-type, consistent with it being an RR Lyr star.  Given its magnitude it really must be a star in our own halo.   It was detected at a very significant level (14$\sigma$)  but the spectrum clearly showed no emission.  The magnitude difference between the WN and CT filter was 0.3~mag.  Our imaging must have caught this star as it was changing brightness most rapidly.

\section{Summary, Discussion, and Future Work}
\label{Sec-End}

We describe here the results of our second year of surveying the Magellanic Clouds for Wolf-Rayet stars.  We have covered 57-60\% of our intended survey area, finding (in total) 13 new WRs in the LMC, bringing the total number of WRs known in the LMC to 152, 8.6\% of which have been found as part of our survey.   What does this imply for the overall numbers?  

In the first year, our survey of the LMC concentrated on the regions where WRs were already known, both as a test of whether or not we would readily detect the known ones (we did, other than in the most crowded regions near R136 in 30 Dor), but also based on the principle that that is where most of the massive stars would be found\footnote{This is perhaps best illustrated by repeating the apocryphal story that when asked why he robbed banks, ``Slick Willie" Sutton replied that it was because that's where the money is.}.  As our result, in our first year of the survey we found nine new WRs in the LMC despite surveying only 15\%.  Thus we were not completely surprised that although we more than doubled the survey area we detected only 4 new WRs this year.  We are encouraged by the fact that we are detecting many Of-type stars with weaker He~II~$\lambda 4686$ than that found in WRs.  We were also encouraged that, despite the large amount of overlap with the regions surveyed in our first year,  we detected nothing new in the regions we had previously surveyed. 
 We have as yet found no new WRs in the SMC, but we can argue that since there are only 12 known WRs in the SMC, a similar 8.6\% increase would  be only 1 star, and thus our not finding any new SMC WRs (yet) may well be due to small number statistics.
 
 Of the 13 LMC WRs we have so far found as part of our survey, 12 of these are WNs.  Although this confirms our suspicion that previous studies have selectively missed WNs because they are weaker lined, it has changed the WC+WO to WN ratio negligibly, from 0.23 (based on the additions and demotions listed by Neugent et al.\ 2012a) to 0.22.  It remains considerably larger than the 0.11 expected from the Geneva evolutionary models computed with $z=0.006$ and an initial rotation of 40\% of the breakup speed (C. Georgy, private communication 2012, and used by Neugent et al.\ 2012b).  What we can say is that this number is less suspect than in the past.  We note that the discrepancy with the models is probably slightly larger than this, as we have included in our counts both the hot and cool ``slash" stars; the evolutionary models can recogize surface abundances, but not the small changes in wind density that would drive H$\beta$ from pure absorption into P Cygni.
 
 Doubtless the most interesting thing revealed by our survey has been the 8 faint WN3 stars which show O3-type absorption spectra, what we are referring to as the WN3/O3s.  These stars are faint visually, with $M_V$$\sim$$-2$ to $-3$, and are much too faint (by 2.5-3 mags!) to actually be WN3+O3~V binaries.  What are these stars?  Our preliminary modeling (Paper~I and Massey et al.\ 2015) shows that it is possible to produce the emission and absorption features with a single set of physical parameters.  These stars appear to be as hot and as bolometrically luminous as other WNs, but have significantly lower mass-loss rates.  Why?  How did these stars evolve?  We believe the answer to this requires a careful comparison of their physical properties with other WNs, and this modeling effort is underway, assisted now by UV data recently obtained with {\it HST}.   We find that one of the stars discovered in the current paper, LMCe069-1, may be the faintest WR star known (in terms of absolute visual magnitude, which is $-1.8$ for this star), and it may also be a little more intermediate in type than the others, showing both stronger emission and weaker absorption.
 
\begin{figure}
\epsscale{1.0}
\plotone{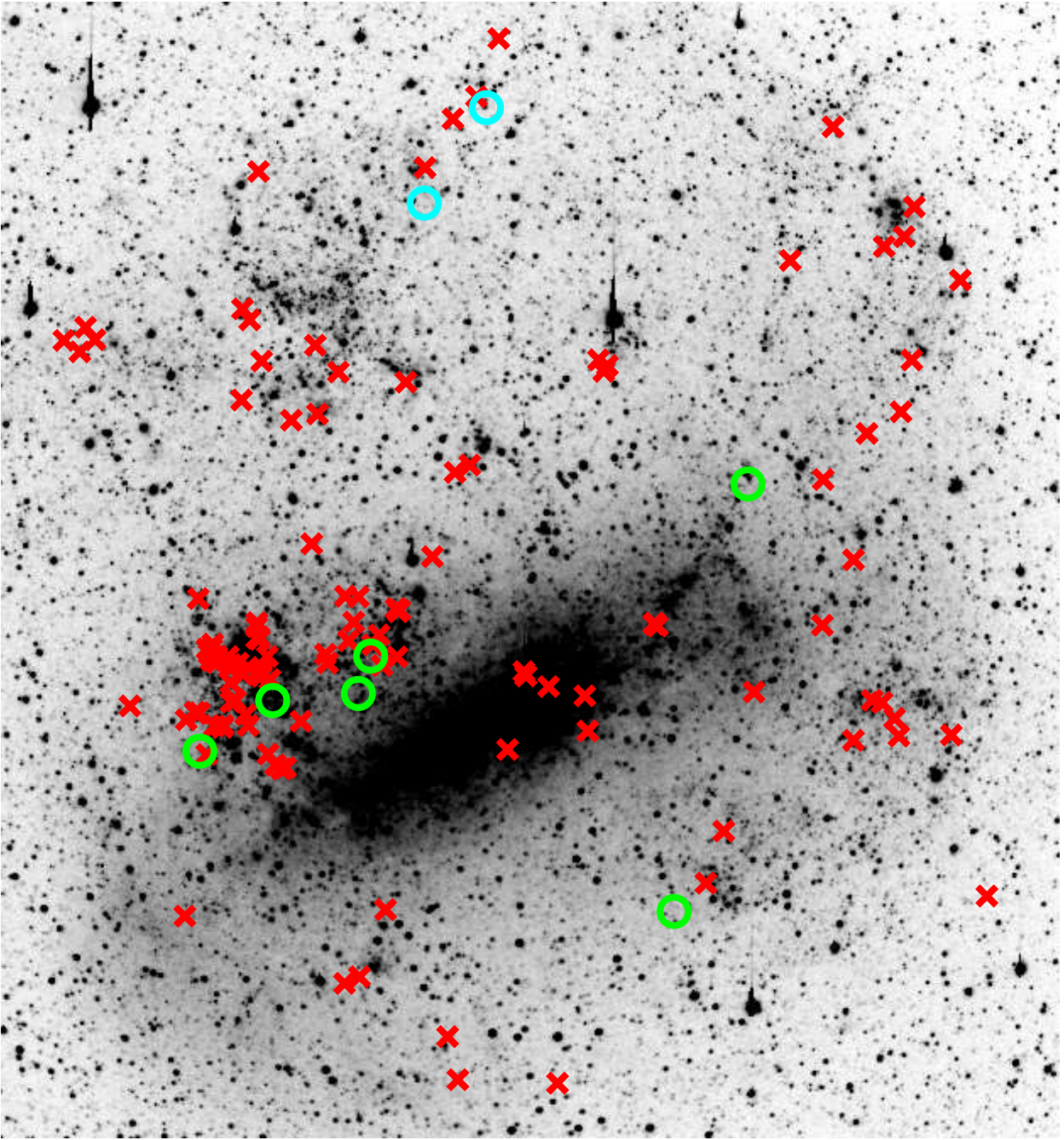}
\caption{\label{fig:LMCWR} The location of WRs in the LMC.  ``Normal" WRs are shown with red x's, while the WN3/O3s are shown as circles (green if from Paper I and cyan from the present paper).}
\end{figure}

We are often asked if there is anything special about the spatial distributions of these WN3/O3 stars in comparison to ``normal" WRs.  If they were separated, or more isolated, one might use this information to infer that they are formed from different progenitors.   Fig.~\ref{fig:LMCWR} shows rather conclusively that there is not.  We have marked the non-WN3/O3s with green x's, and the WN3/O3s with circles (red if found last year, and blue if found this year).  To us, there appears to be nothing unusual in their locations.

We do wish to call attention to one fact that we did not appreciate until recently.  Five of the 12 SMC WRs are also faint ($M_V\sim -3.6$ to $-4.6$), and also show absorption spectra typical of very early O stars (see Table 1 of Massey et al.\  2003).    Perhaps these are similar to the WN3/O3 stars we have discovered in the LMC.  We are currently investigating this possibility.  Although the SMC WRs are often considered to be binaries, we'll note that Conti et al.\ (1989) argued that instead they may simply have weaker stellar winds, allowing the presence of photospheric absorption lines.  This is consistent with the results of our modeling for the LMC WN3/O3s.

We are now at a point in our survey where we feel we should address our completeness.  As discussed in Paper I, surveys such as ours are not simply {\it fluxed-limited}: an Of-type star, for instance, will be bright but with a relatively small ($<$10\AA) equivalent width of He II $\lambda 4686$.  Thus, although there is plenty of flux in the line, the star is hard to detect either photometrically or by image subtraction because the magnitude difference is small.  It is really the {\it contrast} between the brightness in the on-line exposure (through either the {\it WN} or {\it WC} filter) and that in the continuum exposure ({\it CT}) that determines the detectability. The ``flux-limit" argument comes into play in that as we go fainter the photometric errors increase, and thus a given magnitude difference becomes less significance relative to the photometric error.  Note that difference in magnitude between the on-line and off-line filter ($\Delta$ m) is mainly just another way of characterizing the equivalent width.
 
\begin{figure}
\epsscale{1.0}
\plotone{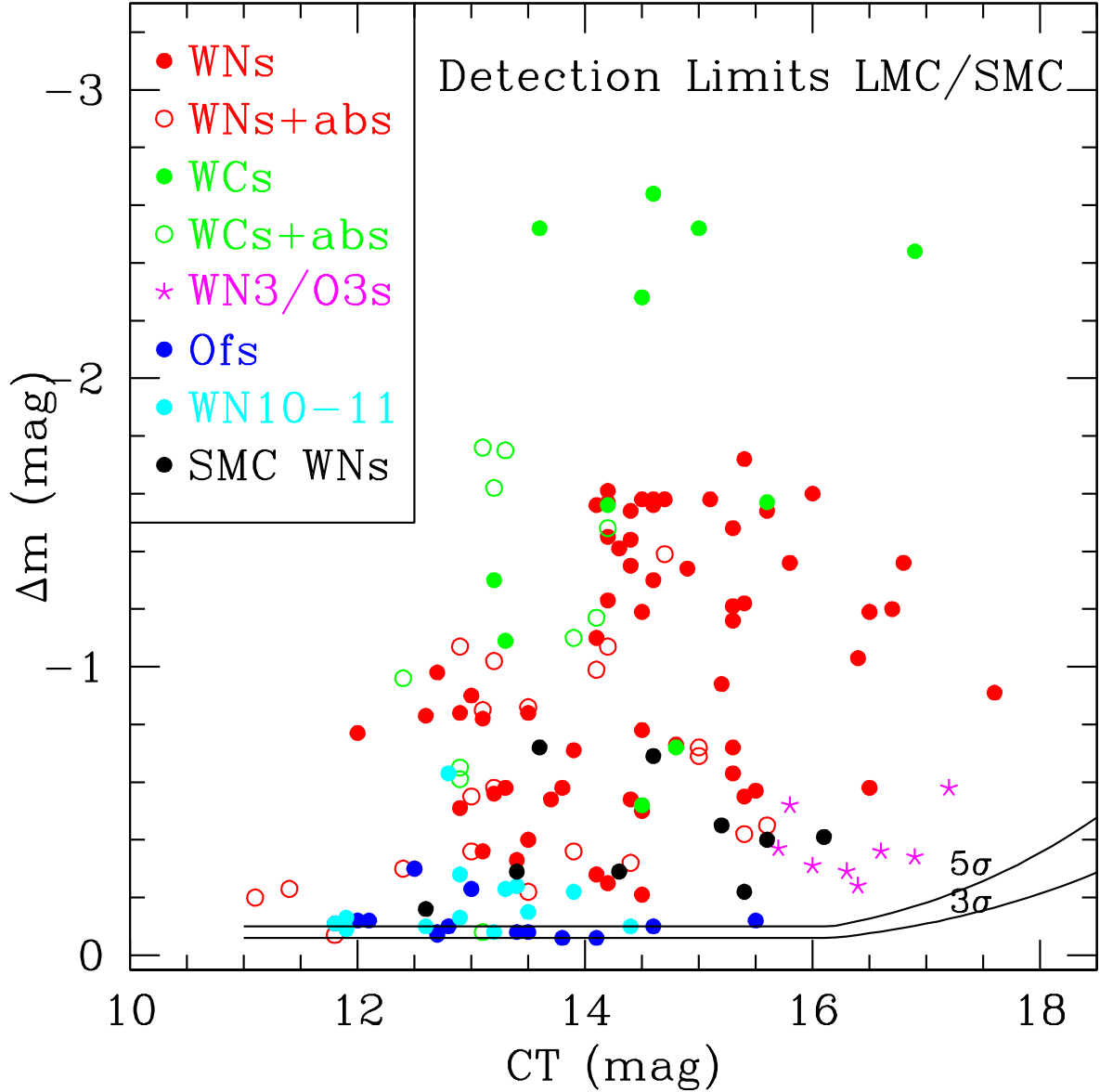}
\caption{\label{fig:complete} Detection limits for our survey.   We plot the magnitude difference $\Delta m$ between the emission-line filter and continuum (whichever was stronger) against the continuum magnitude {\it CT} for both the previously known and newly found WRs in the LMC and SMC.  The WOs are included with the WCs for simplicity.  We have ignored stars in
BAT99 with uncertain or ambiguous spectral types. As expected, the WC stars are easiest to find, and the Of-type and WR ``slash stars" (WN9-10) the hardest.  Our 3$\sigma$ and 5$\sigma$ detection limits are indicated, where we have adopted a conservative 0.02~mag lower limit to the photometric error at the bright end.}
\end{figure}

 Given that we really did not know how deep we needed to go before we discovered the WN3/O3s, it is a legitimate question of {\it are we going deep enough}?  We can answer this by examining the magnitude difference {\it $\Delta m$} against the continuum magnitude {\it CT} in Fig.~\ref{fig:complete}.  Since our survey detected most of the known WRs in our fields (see Paper I) we have constructed this diagram for all of the WRs we found as part of our program (both previously known and newly discovered).  We have also constructed the 3$\sigma$ and 5$\sigma$ error limits in our photometry ({\it WN-CT} or {\it WC-CT}, whichever was stronger).
 There are several noteworthy aspects of this figure.  First, we confirm that the Of-type stars are amongst the hardest to detect.  We also find that the latest type WNs (WN10-11, previously known as the ``slash" WRs) are
 also very difficult to detect, due to their very weak He II $\lambda 4686$ lines.  However, at the same $\Delta m$ that characterizes the WN3/O3s, our 5$\sigma$ detection limit is at least a magnitude fainter than where we are finding these stars.  This strongly suggests that we are not missing a fainter population of WN3/O3s.
  
 We have also used this opportunity to examine our deep WR surveys of M33 (Neugent \& Massey 2011) and M31 (Neugent et al.\ 2012b).   Our
 conclusions are that while we detected the vast majority of WRs (if they are similar to those in the LMC), we certainly would have missed the WN3/O3s in these galaxies: the combination of faintness with weak lines would have placed them below our detection limits in those galaxies.  Thus, we plan a deeper survey of these galaxies in the near future.
  
As mentioned in Section~\ref{Sec-intro}, we expect to finish our survey during the next Magellanic Cloud season, weather permitting.  When we are done, we will have a thorough census of the Wolf-Rayet content in these nearby, relatively metal-poor galaxies.  While we expected to emerge with a solid knowledge of the numbers and types of WRs in these galaxies, the real value has been in what we did not expect to find, such as the WN3/O3 and various peculiar emission-line stars described here and in Paper I.   We look forward to the new discoveries that await us next year.

\acknowledgements
We are grateful for the excellent support we always receive at Las Campanas Observatory, as well as the generosity of the Carnegie Observatory and Steward Observatory Arizona Time Allocation Committees.  We are also indebted to the anonymous referee for useful suggestions, which have improved the paper. We thank Carlos Contreras for his helpful advice on the e2v camera, and Nolan Walborn, John Hillier, and Howard Bond for useful comments.  Support for this project was provided by the National Science Foundation through AST-1008020, and through Lowell Observatory.   This research has made use of the VizieR catalogue access tool, CDS, Strasbourg, France. The original description of the VizieR service was  published in A\&AS 143, 23.   We note in particular the usefulness of the Catalog of Stellar Spectral Classification prepared by our colleague at Lowell Observatory, Brian Skiff.   We also made use of data products from the Two Micron All Sky Survey (2MASS),  which is a joint project of the University of Massachusetts and the Infrared Processing and Analysis Center/California Institute of Technology, funded by the National Aeronautics and Space Administration and the NSF. 

{\it Facilities:} \facility{Magellan:Clay(MagE spectrograph)}, \facility{Swope (e2v imaging CCD)}

%\clearpage

%\clearpage

\clearpage

\begin{deluxetable}{l c c c c c c c c c c c l  l}
\rotate
\tabletypesize{\scriptsize}
\tablecaption{\label{tab:WRs} Newly Found WRs}
\tablewidth{0pt}
\tablehead{
\colhead{ID\tablenotemark{a}}
& \colhead{$\alpha_{\rm 2000}$} 
& \colhead{$\delta_{\rm 2000}$} 
& \colhead{$V$\tablenotemark{b}}
&\colhead{$B-V$\tablenotemark{b}}
& \colhead{{\it CT}} 
& \multicolumn{2}{c}{\it WN - CT}
&
& \multicolumn{2}{c}{He~II $\lambda 4686$} 
& \colhead{$M_V$\tablenotemark{c}}
& \colhead{Sp.\ Type} 
& \colhead{Comment} \\  \cline{7-8} \cline{10-11}
&&&&&
&\colhead{mag} 
& \colhead{$\sigma$}
& \colhead{}  
& \colhead{$\log$(-EW)} 
& \colhead{FWHM} 
}
\startdata   
LMCe063-1& 05 38 24.30 & $-69$ 29 13.5  & 13.04\tablenotemark{d} & +0.28\tablenotemark{d} & 13.2 & $-0.08$ & 4.1 && 0.4   & 8 & $-5.9$ & WN11 & Sk$-69^\circ$240 \\
LMCe132-1 &  05 14 17.55 & $-67$ 20 35.7 &  14.34 & $-0.13$ &14.4 & -0.08 & 4.2 &&  0.7  & 8   & $-5.9$ & O3.5~If*/WN5  \\
LMCe159-1 & 05 24 56.89 & $-66$ 26 44.5 & 16.34 & $-0.23$ & 16.4 & -0.22 & 10.2 && 1.3 & 20 & -2.6 & WN3/O3 \\
LMCe169-1 & 05 21 22.84 & $-65$ 52 49.0 & 17.12 & $-0.19$ & 16.9 & -0.34 & 15.5 && 1.4 & 27 & -1.8 & WN3/O3 \\
\enddata
\tablenotetext{a}{Designation from the current survey.  We have denoted the e2v fields with a small ``e" to distinguish them from our numbering system from Paper I, i.e., LMCe159 is distinct from LMC159.  We plan to impose less idiosyncratic designations once our survey is complete.}
\tablenotetext{b}{Photometry from Zaritsky et al.\ 2004.}
\tablenotetext{c}{We assume an apparent distance modulus of 18.9 for the LMC, corresponding to a distance of 50 kpc (van den Bergh 2000) and an average extinction of $A_V=0.40$ (Massey et al.\ 1995, 2007).}
\tablenotetext{d}{Based upon our spectrophotometry, the ``emission  free" values would be $V\sim13.11$ and $B-V\sim+0.23$, derived from the continuum fluxes at 4400\AA\ and 5500\AA.}
\end{deluxetable}

\begin{deluxetable}{l c c c c c c c l l}
%\rotate
\tabletypesize{\scriptsize}
\tablecaption{\label{tab:others} Other Emission-Line Stars}
\tablewidth{0pt}
\tablehead{
\colhead{ID\tablenotemark{a}}
& \colhead{$\alpha_{\rm 2000}$} 
& \colhead{$\delta_{\rm 2000}$} 
& \colhead{$V$\tablenotemark{b}}
&\colhead{$B-V$\tablenotemark{b}}
& \multicolumn{2}{c}{He~II $\lambda 4686$} 
& \colhead{$M_V$\tablenotemark{c}}
& \colhead{Sp.\ Type} 
& \colhead{Comment} \\  \cline{6-7} 
&&
& \colhead{}  
& & \colhead{$\log$(-EW)} 
& \colhead{FWHM} 
}
\startdata   
LMC156-2  & 04 51 10.59 & $-69$ 33 21.1 & 13.65 & $-0.13$ & \nodata & \nodata & $-5.3$ & Onfp &  \\
LMCe034-1 & 05 31 42.05 & $-70$ 37 54.0  & 18.02\tablenotemark{d} & -0.09\tablenotemark{d} & 1.8  & \nodata & $-2.2$ & \nodata & NE member of 2\farcs1 pair. \\
\enddata
\tablenotetext{a}{Designation from the current survey.  We have denoted the e2v fields with a small ``e" to distinguish them from our numbering system from Paper I, i.e., LMCe159 is distinct from LMC159.  We plan to impose less idiosyncratic designations once our survey is complete.}
\tablenotetext{b}{Photometry from Zaritsky et al.\ 2004 unless otherwise noted.}
\tablenotetext{c}{We assume an apparent distance modulus of 18.9 for the LMC, corresponding to a distance of 50 kpc (van den Bergh 2000) and an average extinction of $A_V=0.40$ (Massey et al.\ 1995, 2007).}
\tablenotetext{d}{``Emission free" photometry derived from the continuum fluxes at 4400\AA\ and 5500\AA.}
\end{deluxetable}

%\clearpage

\thispagestyle{empty}
\begin{deluxetable}{l c c c c c l }
%\rotate
\tabletypesize{\scriptsize}
\tablecaption{\label{tab:losers} Non Emission-Line Stars}
\tablewidth{0pt}
\tablehead{
\colhead{ID\tablenotemark{a}}
& \colhead{$\alpha_{\rm 2000}$} 
& \colhead{$\delta_{\rm 2000}$} 
& \colhead{$V$\tablenotemark{b}}
&\colhead{$B-V$\tablenotemark{b}}
&\colhead{$M_V$\tablenotemark{c}}
&\colhead{Type} \\
}  
\startdata
SMCe107-1 & 01 05 55.19 & $-71$ 15 48.2 & 15.45 & $-0.23$ &$-3.7$ &B1-B1.5 III\\
SMCe107-2 & 01 06 13.72 & $-71$ 11 22.3  & 15.06\tablenotemark{d} & -0.12\tablenotemark{d} & $-4.0$& B2 I\\
LMC197-1   & 05 23 41.11 & $-69$ 04 45.4  &  16.54 & $-0.12$ & $-2.4$ &  B3 III\\
LMC222-3   & 05 12 54.99 & $-68$ 41 12.0 &   15.82 & $+0.08$ & $-3.1$ & B1 V\\
LMCe019-1 & 05 11 17.41 & $-71$ 02 44.2 &  16.21\tablenotemark{e} & \nodata & \nodata &RR Lyr (A-type)\\
LMCe050-1 & 05 43 58.62 & $-70$ 14 21.6 &  13.93 & $-0.21$ & $-5.0$&B2.5 I \\
LMCe116-1  & 05 00 12.31 & $-67$ 45 17.5 & 14.89 & $-0.02$ & $-4.0$ &B2 III \\
\enddata
\tablenotetext{a}{Designation from the current survey.  We have denoted the e2v fields with a small ``e" to distinguish them from our numbering system from Paper I, i.e., LMCe159 is distinct from LMC159.  We plan to impose less idiosyncratic designations once our survey is complete.}
\tablenotetext{b}{Photometry from Zaritsky et al.\ 2002 for the SMC stars, and Zaritsky et al.\ 2004 for the LMC stars, unless otherwise noted.}
\tablenotetext{c}{We assume apparent distance moduli of 19.1 and 18.9 for the SMC and LMC respectively, corresponding to distances of 59~kpc and 50 kpc (van den Bergh 2000), respectively, and average  extinctions of $A_V=0.28$ and $A_V=0.40$ (Massey et al.\ 1995, 2007).}
\tablenotetext{d}{From Zacharias et al.\ 2013.}
\tablenotetext{e}{From Soszy\'{n}ski et al.\ 2009.}
\end{deluxetable}
\end{document}